\documentclass[aps,prb,superscriptaddress, twocolumn, showpacs]{revtex4}
\usepackage{amsmath}
\usepackage{amssymb}
\usepackage{graphicx}
\usepackage{color}
\usepackage{tikz}
\usepackage{pgffor}
\usepackage{verbatim}
\usepackage[normalem]{ulem}
\usepackage{soul}
\begin{document}

\title{Fluctuation relations for spin currents}

\author{Chenjie Wang}
\affiliation{James Franck Institute and Department of Physics, University of Chicago, Chicago, Illinois 60637, USA}

\author{D. E. Feldman}
\affiliation{Department of Physics, Brown University, Providence, Rhode Island, 02912, USA}
\date{\today}

\begin{abstract}
The fluctuation theorem establishes general relations between transport coefficients and fluctuations in nonequilibrium systems. Recently there was much interest in quantum fluctuation relations for electric currents. Since charge carriers also carry spin, it is important to extend the fluctuation theorem to spin currents. We use the principle of microscopic reversibility to derive such theorem. As a consequence, we obtain a family of relations between transport coefficients and fluctuations of spin currents. We apply the relations to the spin Seebeck effect and rectification of spin currents.  Our relations do not depend on a microscopic model and hence can be used to test the validity of theoretical approximations in spin-transport problems.
\end{abstract}


\maketitle

\section{Introduction}

Both charge and spin of electron play major roles in solid state physics and its applications but the electric and spin currents behave in different ways. Coulomb interaction brings a high energy cost to any violation of charge neutrality. This fact and charge conservation make it easy to create and manipulate charge currents. At the same time, spin does not conserve. Its coupling to external probes is weaker than that of the electric charge. Thus, manipulating spin currents is a harder task than working with electric currents. Recently, much effort has focused on overcoming this challenge in the hope to get more control over the spin degree of freedom
\cite{spintronics}. That effort has been motivated by both potential applications and the interest in basic spin physics. An important open problem in the field is the understanding of nonlinear spin transport.

The problem is challenging because nonlinear transport occurs away from thermal equilibrium. Until recently, any progress in the theory of nonequilibrium transport was inevitably based on microscopic models
\cite{kynetics}. The situation changed after the discovery of fluctuation theorems \cite{BK1,BK2,BK3,fl0,fl1,fl2} which brought a powerful general principle to nonequilibrium statistical mechanics. The goal of this paper consists in the derivation of a family of fluctuation relations for spin currents and noises.
We then apply them to several spin-transport problems, including rectification of spin currents and the spin Seebeck effect \cite{see1,see2,see3}, i.e., the generation of a spin current by a temperature gradient.  An important application is the theory of spintronic nanomachines\cite{utsumi15}.
Our fluctuation relations are completely general and do not depend on microscopic details or the choice of a model.
This makes their application particularly interesting.
They also give a useful tool for testing theoretical approximations.

The history of fluctuation relations is fascinating. They were first introduced by Bochkov and Kuzovlev in the late 1970s \cite{BK1,BK2,BK3}. The name ``fluctuation theorem'' was coined by H\"anggi in 1978 \cite{name}.  Yet, its importance has been fully appreciated only after a series of seminal results in the 1990s, in particular, the works by Evans, Cohen and Morriss\cite{Cft}, Jarzinski\cite{Jft}, and Crooks\cite{Crft}.
Fluctuation theorems generalize the principle of detailed balance. Detailed balance establishes a fundamental relation between the rates of different processes. Fluctuation theorems go a step further and relate the probabilities of a process and its time-reversed version on a macroscopic time scale. At first sight, fluctuation relations apply to time-reversal invariant systems only. Indeed, a process and its time-reversed version can occur in the same system only if the time-reversal symmetry is present, i.e., in the absence of a spontaneous magnetization and external magnetic fields. Otherwise, the time-reversed process would require changing the sign of the magnetic field. However, interesting relations do not have to connect the properties of a single system.
This fact was already recognized in the pioneering papers \cite{BK2,BK3}.
A number of fluctuation relations for electric currents in a magnetic field were derived in Ref. \onlinecite{saito}. Those relations contain currents and noises in two systems that differ by their directions of the magnetic fields. In some cases it even proved possible to derive fluctuation relations that connect currents and noises in a single system in the absence of the time-reversal symmetry \cite{wang1,wang2}. A review of those developments can be found in Ref. \onlinecite{wang3}.

Focusing on the spin degree of freedom means treating currents as multi-component with the components, corresponding to the up and down spin projections. A fluctuation  theorem for multi-component currents was addressed in Ref. \onlinecite{spin-utsumi}. That theorem is not directly relevant in our problem since Ref. \onlinecite{spin-utsumi} considers the currents of time-reversal-invariant quantities while spin changes its sign under time reversal. More specifically, the spin operator $\mathbf S$ satisfies
\begin{equation}
\label{0}
\Theta \mathbf S\Theta^{-1} = -\mathbf S,
\end{equation}
where $\Theta$ is the time reversal operator. In contrast, charge is invariant under time reversal: $\Theta Q\Theta^{-1}=Q$. Several fluctuation relations for spin currents were derived in Ref. \onlinecite{lopez12}. In this paper we establish a bigger family of relations which could not be obtained with the approach of Ref. \onlinecite{lopez12}. Indeed, Ref. \onlinecite{lopez12} did not make use of microreversibility that allows us to go beyond the relations \cite{lopez12}.

According to the principle of microscopic reversibility, the amplitudes of a process in a given system and the time-reversed process in the system with the reversed directions of all magnetic fields must be complex conjugate. On the basis of calculations within the Landauer-B\"uttiker formalism \cite{forster08,forster09}, it was proposed that microreversibility fails in nonzero magnetic fields. We believe that the conflict of the Landauer-B\"uttiker approximation and the principle of microscopic reversibility shows only the limitations of the Landauer-B\"uttiker approach. Note that the conclusions, based on the microreversibility principle are supported by the experiment
\cite{exp1,exp2} and agree with theoretical calculations beyond the Landauer-B\"uttiker approximation \cite{saito09,nasb10}. A detailed discussion of microreversibility in the presence of a magnetic field can be found in the review \cite{wang3}. To make this paper self-contained we repeat some of that discussion in Appendix \ref{sec:microreversibility}. Clearly, it is important to address the consequences of microreversibility for spin currents and noises.

Besides Refs. \onlinecite{saito,spin-utsumi,lopez12}, several other articles address related physics. In particular, Refs. \onlinecite{Aft,Bft} explore fluctuation relations for time-dependent protocols that can break time-reversal symmetry. In contrast to our work and similar to Ref.~\onlinecite{spin-utsumi}, they do not consider observables that change their sign under time-reversal.

The connection of our results with the pioneering work \onlinecite{BK1,BK2,BK3} is particularly interesting since these works consider quantities, even and odd under time-reversal, on equal footing. Refs. \onlinecite{BK1,BK2,BK3} contain great wealth of results. In particular, Ref. \onlinecite{BK1} derives general fluctuation relations in thermal equilibrium in the absence of magnetic fields. Our results are related to those of Ref. \onlinecite{BK1} in the equilibrium limit at zero magnetic field. At the same time, our main focus is on nonequilibrium transport, driven by the differences of the reservoir temperatures and chemical potentials, in an external magnetic field.

Refs. \onlinecite{BK2,BK3} derive fluctuation relations for nonequilibrium steady states that form after a sudden change of parameters in a system which is initially in equilibrium and does not break time-reversal symmetry
(see Section 7 of Ref. \onlinecite{BK3}). This problem is related to but different from ours. Indeed, the meaning of the fluctuation relations \cite{BK3} is different from our work. Ref. \onlinecite{BK3}
allows one to connect the derivatives of various correlation functions with respect to the parameters of the Hamiltonian. This is relevant in the problem of quantum quench (see, e.g., Ref. \onlinecite{quench}).
At the same time, our focus is on the dependence of correlation and response functions on the reservoir temperatures and chemical potentials. This is a natural question in the context of mesoscopic transport \cite{datta}.
Note that the approach \cite{BK3} has been used to derive fluctuation relations for electric currents, driven by a voltage bias between the reservoirs of the same temperature. This proved possible because electro-neutrality dictates that a change of the potential energy of a reservoir leads to an equal change of its electro-chemical potential. It is unclear if a similar approach can be used in our problem which involves gradients of the temperature and of the chemical potential, conjugate to the spin. We use a different method that builds on Refs.  \onlinecite{andrieux09,campisi10}.

Our paper has the following structure. We set the problem and derive a general fluctuation theorem for spin transport in Section \ref{ft}.  In the third Section we deduce from the theorem a family of fluctuation relations for spin currents and noises.
Section \ref{sec:example} applies our relations to the problems of spin-current rectification and to multi-terminal spintronic devices. We compare our results with the previous work in subsection \ref{sec:comparison}. Section \ref{sec:thermo} focuses on thermospin transport and Section \ref{sec:con} summarizes our results. In Appendix \ref{sec:microreversibility} we discuss the microreversibility controversy. Appendix \ref{sec:spinchemopoten} reviews ways to measure spin currents and chemical potentials. Technical details of our calculations are relegated to  Appendices \ref{sec:ho} and \ref{sec:IV}. Appendix \ref{sec:chiral} addresses spin transport in chiral system, where stronger and simpler fluctuation relations can be derived.

\section{Fluctuation theorem}
\label{ft}

\begin{figure}[b]
\includegraphics[width=2.5in]{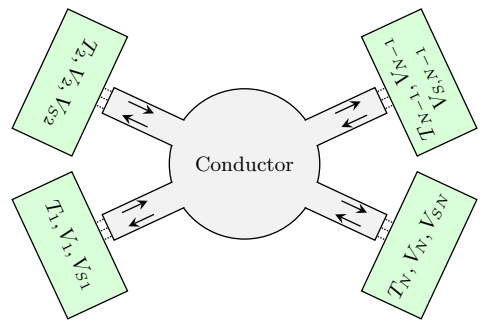}
\caption{Sketch of the $N$-terminal setup (only four terminals are shown) for spin, charge and heat transport.}\label{fig_multi}
\end{figure}

{We will study fluctuation theorems that} are concerned with the transport through a finite system (conductor) attached to two or more large reservoirs (Fig.~\ref{fig_multi}). Each reservoir is assumed to be in thermal equilibrium but the reservoirs are not necessarily in equilibrium with each other. A difference between the reservoirs' chemical potentials drives a current through the conductor. {Our goal is to establish various relations between the correlation and response functions of the currents.} The crucial assumption is that the system does not have long-term memory beyond a characteristic transport time through the conductor. Thus, we can assume any convenient choice of the system's past. The standard choice is as follows. Initially, the reservoirs and the conductor are disconnected. Next, the particle and energy exchange between the system and reservoirs is turned on for some long time $\tau$. At the end, the reservoirs are again disconnected from the conductor. We are interested in the transport during the time interval $\tau$. Yet, it is most convenient to describe such transport in terms of the quantities, measured before the interaction between the conductor and the reservoirs was turned on at $t=0$ and after the interaction
was turned off at $t=\tau$. For example, the  average current from reservoir number $k$ is defined as $-[Q_k(\tau)-Q_k(0)]/\tau$, where $Q_k(t)$ is the  {total} charge of reservoir $k$ at time $t$. The energy and spin currents are defined in a similar way. Thus, the following measurement protocol is assumed: we measure the charge, the $z$-projection of the spin and the energy of each reservoir before and after the evolution period $\tau$.

Such simultaneous measurements are only possible, if the charge and the $z$-component of the spin commute with the energy operator and hence conserve. This poses an obvious problem: spin does not conserve because of the spin-orbit interaction. In addition, external magnetic fields may lead to the Larmor precession which also breaks spin conservation. Moreover, if the spin-relaxation time is short it may be meaningless to even speak about spin currents, flowing from the reservoirs. Indeed, the spin current is defined as the total spin flowing per unit time through a cross-section
near the boundary of the reservoir and conductor. In the absence of spin conservation such current depends on the choice of the cross-section.

The above discussion necessitates the following assumptions. First, to avoid the problem with the Larmor precession, we assume that the magnetic fields are oriented along the $z-$axis in the reservoirs. Such magnetic fields would break the conservation of the $x$- and $y$-components of the reservoir spin but we are only interested in the $z$-component. Second, we assume a weak spin-orbit interaction in the reservoirs. The interaction must be weak enough that it can be neglected on the time scale of the transport through the conductor and on the much longer time scale during which a quasiequilibrium establishes in a large portion of each reservoir near the interface with the conductor.
Such quasiequilibrium can be described by two different chemical potentials in each reservoir for the up and down projections of the electron spin on the $z$-axis. In the simplest situation the two chemical potentials are equal but we will consider the most general case of arbitrary chemical potentials. We will assume that the quasiequilibrium survives during the evolution period $\tau$. Note that very long spin-relaxation times were reported for donor
spins in pure silicon
\cite{silicon}.

The reservoir spin is gradually lost due to the spin-orbit interaction and the spin current into the conductor. Hence, the two chemical potentials slowly depend on time. To keep them constant, one would need to compensate the spin loss by a spin current injection into the reservoir.

We do not make any assumptions about the spin-orbit interaction and the direction of the magnetic field in the conductor. If the $z$-projection of the spin does not conserve in the conductor then the spin currents, leaving the reservoirs, do not add up to zero.

We are now ready to derive the fluctuation theorem. Our method is connected with the approach of Refs. \onlinecite{andrieux09,campisi10}. We assume that initially all reservoirs are disconnected. It will be convenient to include the conductor as a part of one of the reservoirs. This requires us to make the Hamiltonian of the conductor time-dependent so that the spin conserves in the conductor at $t<0$ and $t>\tau$. We will assume that the time-dependence is slow (i.e., $\tau$ is large) and hence the energy conserves. We will use the units such that $\hbar=k_B=1$. We also assume in this section that the temperature is the same in all reservoirs. Spin currents, driven by temperature gradients, will be addressed in Section V.

Specifically, we assume the following protocol. The system is described by the time-dependent Hamiltonian
$H=\tilde H_0(t)+\sum_{i=1}^N(\tilde H_i+\tilde H_{0i}(t))$,
where $N$ is the number of the reservoirs, $\tilde H_i$ are time-independent Hamiltonians of the reservoirs, $\tilde H_0(t)$ is the Hamiltonian of the conductor, and $\tilde H_{0i}(t)$ describes the interaction of the conductor with reservoir $i$. We asuume $\tilde H_{0i}(t)=0$ at $t<0$ and $t>\tau$. Also, $\tilde H_{0i}(t)$ are time-independent at times $t$ such that $\tau_0<t<\tau-\tau_0$, where $\tau_0\ll\tau$ is the interaction-switching time.  $\tau_0$ should be large enough so that we can employ energy conservation. We also assume that $\tilde H_0$ does not depend on time at $t<0$ and $t>\tau$, and $\tilde H_0(t<0)=\tilde H_0(t>\tau)$. Its precise form is unimportant at such $t$ but it is assumed that at $t<0$ and $t>\tau$ the $z$-component of the spin conserves in the conductor, i.e., the spin-orbit interaction is absent. At $\tau_0<t<\tau-\tau_0$, $\tilde H_0$ is time-independent and equals the experimentally relevant Hamiltonian of the conductor. In what follows we absorb all parts of the system into various reservoirs and thus redefine the reservoir Hamitonians as $H_1(t)=\tilde H_1+\tilde H_{01}(t)+\tilde H_0(t)$ and $H_k(t)=\tilde H_k+\tilde H_{0k}(t)$, $k>1$. Each reservoir is initially ($t=0$) in equilibrium. Different reservoirs are not in equilibrium with each other, i.e., their chemical potentials are not the same. Since the conductor is absorbed into the first reservoir, its initial state is equilibrium with the same chemical potentials as in the first reservoir. Certainly, the finite size of the conductor means that the choices of its initial state and initial Hamiltonian do not affect our results.

The initial density matrix of the system is the product of the density matrices of the reservoirs:
\begin{equation}
\label{1}
\rho_F=\prod_i\rho_i=\prod_i\frac{1}{Z_i}\exp\left[\frac{V_iQ_i+V_{Si}S_{zi}-E_i}{T}\right],
\end{equation}
where
\begin{equation}
\label{2}
Z_i={\rm Tr}\exp\left[\frac{V_i Q_i+V_{Si}S_{zi}-E_i}{T}\right],
\end{equation}
$Q_i$ is the charge of reservoir $i$, $E_i$ is its energy, $S_{zi}$ is the $z$-component of the spin in reservoir $i$, $V_i$ is the electro-chemical potential and $V_{Si}$ is the spin chemical potential, conjugate to $S_{zi}$
(experimental procedures for the detection of spin currents and spin chemical potentials are discussed in Appendix \ref{sec:spinchemopoten}). The chemical potentials for the up and down spins are $(-e)V_i\pm V_{Si}/2$ and equal $-eV_i$ at $V_{Si}=0$, where
$-e$ is the electron charge. Eq. (\ref{1}) assumes that all reservoirs have the same temperature $T$.


The evolution operator is
\begin{equation}
\label{3}
U_F(t)={\rm \hat T}\exp\left(-i\int_0^t H(t')dt'\right)
\end{equation}
where $H(t)=\sum H_i(t)$ is the time-dependent Hamiltonian, {and $\rm\hat T$ is the time ordering operator.}

We can now write the probability of the {\it forward} process. This is the probability that during the evolution period $\tau$ the charges of the reservoirs change by $\Delta Q_i$ and their spins change by $\Delta S_{zi}$:
\begin{widetext}
\begin{align}
\label{4}
P_F(\Delta {\bf Q}, \Delta {\bf S}_{z})=\sum & \langle{\bf Q}(0), {\bf S}_{z}(0),n(0)|\rho_F|{\bf Q}(0), {\bf S}_{z}(0),n(0)\rangle \times \big|\langle {\bf Q}(\tau), {\bf S}_{z}(\tau),n(\tau)|U_F(\tau)|{\bf Q}(0), {\bf S}_{z}(0),n(0)\rangle\big|^2  \nonumber\\
 & \times\delta(\Delta {\bf Q}-({\bf Q}(\tau)-{\bf Q}(0)))
\delta (\Delta {\bf S}_{z}-({\bf S}_{z}(\tau)-{\bf S}_{z}(0))),
\end{align}
\end{widetext}
where ${\bf A}$ stays for a vector whose component $A_i$ describes reservoir $i$. The summation in Eq. (\ref{4}) extends over all initial and final states which we label with the charge vector ${\bf Q}$, the spin vector ${\bf S}_z$
and an additional quantum number $n$ that distinguishes quantum states with identical charges and spins in the reservoirs.

We now introduce the time-reversed {\it backward} process. The initial density matrix
\begin{equation}
\label{5}
\rho_B=\Theta \rho_F\Theta^{-1}=\prod_i\frac{1}{Z_i}\exp\left[\frac{V_iQ_i-V_{Si}S_{zi}-E_i}{T}\right],
\end{equation}
where $\Theta$ is the antiunitary time-reversal operator, satisfying $\Theta i=-i\Theta$, and we use the relation $\Theta S_{zi}\Theta^{-1}=-S_{zi}$. The dynamics is controlled by the Hamiltonian
\begin{equation}
\label{6}
H_B=\Theta H(\tau-t)\Theta^{-1}.
\end{equation}
The evolution operator
\begin{align}
\label{7}
U_B(t)= & {{\rm \hat T}\exp\left(-i\int_0^t H_B(t')dt'\right)}\nonumber \\
=& { \Theta \left[
{\rm \hat T^{-1}}\exp(+i\int_{\tau-t}^\tau  H(t') dt')\right]\Theta^{-1},}
\end{align}
where $\rm \hat T^{-1}$ is the reversed time ordering operator. Hence,
\begin{equation}
\label{8}
U_B(\tau)=\Theta U^\dagger_F(\tau)\Theta^{-1}.
\end{equation}
This equation expresses the microreversibility principle.

We compute the probability that during the evolution period $\tau$ the changes of the reservoir charges are given by the vector $-\Delta{\bf Q}$ and the reservoir spins change by $+\Delta {\bf S}_z$:
\begin{widetext}
\begin{align}
\label{9}
 & P_B(-\Delta {\bf Q}, \Delta {\bf S}_{z})=  \sum  [\langle{\bf Q}(\tau), {\bf S}_{z}(\tau),n(\tau)|\Theta^{-1}]\rho_B[\Theta|{\bf Q}(\tau), {\bf S}_{z}(\tau),n(\tau)\rangle] \nonumber\\ &
\times\big|[\langle {\bf Q}(0), {\bf S}_{z}(0),n(0)|\Theta^{-1}]U_B(\tau)[\Theta|{\bf Q}(\tau), {\bf S}_{z}(\tau),n(\tau)\rangle]\big|^2 \delta(-\Delta {\bf Q}-({\bf Q}(0)-{\bf Q}(\tau)))
\delta (\Delta {\bf S}_{z}-({\bf S}_{z}(0)-{\bf S}_{z}(\tau))).
\end{align}
The above formula differs from Eq. (\ref{4}) in that $\tau$ stays in place of $0$ and {\it vice versa} inside the bra and ket vectors. This change of notation emphasizes the time-reversed dynamics. For the same reason, all matrix elements are written in the basis of the time-reversed states $\Theta|{\bf Q}, {\bf S}_z, n\rangle$. Next, we combine Eqs. (\ref{4}), (\ref{9}), (\ref{1}), (\ref{5}) and (\ref{8}) and use the conservation of the
energy $\sum E_i$ on the evolution period $\tau$. We find
\begin{equation}
\label{10}
P_F(\Delta{\bf Q},{\Delta{\bf S}_{z}})=P_B(-\Delta{\bf Q},{\Delta{\bf S}_z})\exp\left[-\frac{\sum(V_i\Delta Q_i+V_{Si}\Delta S_{zi})}{T}\right].
\end{equation}
Finally, we note that $P_B$ can be interpreted as the probability of a forward process in the system with the same chemical potentials ${\bf V}$ and the opposite chemical potentials $-{\bf V}_S$ in the opposite magnetic field. Here, the fact that ${\bf V}_S$ changes its sign follows from the comparison of Eqs. (\ref{1}) and (\ref{5}). Strictly speaking, in addition to the sign changes of the magnetic field and ${\bf V}_S$, one also needs to demand the opposite sign of the reservoir magnetizations in a system with spontaneously broken time-reversal symmetry. In what follows we instead assume that the time-reversal symmetry is always broken explicitly by some (possibly infinitesimal) magnetic field. Interpreting Eq. (\ref{10}) as a relation between the probabilities of the two forward processes and getting rid of the indices $B$ and $F$, one obtains the {\it fluctuation theorem}
\begin{equation}
\label{11}
P(\Delta{\bf Q},{\Delta{\bf S}_{z}}, {\bf V}, {\bf V}_S, {\bf B})= P(-\Delta{\bf Q},\Delta{\bf S}_z, {\bf V}, -{\bf V}_S, -{\bf B})
\exp\left[-\frac{\sum(V_i\Delta Q_i+V_{Si}\Delta S_{zi})}{T}\right],
\end{equation}
\end{widetext}
where ${\bf B}$ is the magnetic field. The conservation of $S_z$ in the reservoirs implies that the magnetic field is uniform inside each reservoir. Otherwise, the Maxwell equations would require the field to have a nonzero $x$ or $y$ component. The field does not have to be uniform in the conductor.

It what follows it will be convenient to slightly modify the notations. We will redefine $V_i\rightarrow V_i-V$, where $V$ is the common electro-chemical potential of all reservoirs in equilibrium.  This will not change the form of Eq. (\ref{11}) but will allow us to assume that $V_i$ are small in the limit of the low voltage bias. { On the other hand, $V_{Si}$ are already small in the limit of the low voltage bias since spin is not conserved
far from the conductor and hence its chemical potential is zero in a true thermal equilibrium.}

\section{Fluctuation relations for transport coefficients}
\label{sec:fr}

The fluctuation theorem (\ref{11}) is a relation for the probability distribution functions which are hard to extract from experiment. Transport data, such as electric and spin currents,  conductances and noises are among the most important observables. In this section, we establish several relations ({\it fluctuation relations}) among these transport quantities from the fluctuation theorem (\ref{11}).

\subsection{Definitions of transport quantities}
\label{sec:definition}

Let us begin with the definitions of the transport parameters. Recall that our system remains in a steady state for most of the protocol duration $\tau$ (Sec.~\ref{ft}). Under this assumption, the average electric current $I_i$ and the average spin current $I_{Si}$, injected into the conductor from reservoir $i$, can be written as
\begin{align}
\label{13}
I_i &=-\lim_{\tau \rightarrow \infty}\frac{1}{\tau}\langle \Delta Q_i \rangle_{\mathbf B, \mathbf V, \mathbf V_S},\\
\label{14}
I_{Si} &=-\lim_{\tau \rightarrow \infty} \frac{1}{\tau}\langle \Delta S_{zi}\rangle_{\mathbf B, \mathbf V, \mathbf V_S},
\end{align}
where the brackets $\langle \dots \rangle_{\mathbf B, \mathbf V, \mathbf V_S}$ mean averaging over the distribution function $P(\Delta \mathbf Q, \Delta \mathbf S_{z}, \mathbf V, \mathbf V_S, \mathbf B)$. When no ambiguity arises, we will simply use $\langle \dots\rangle$ instead of $\langle \dots \rangle_{\mathbf B, \mathbf V, \mathbf V_S}$ for brevity. In the limit of low voltages, one can expand $I_i$ and $I_{Si}$ as
\begin{align}
I_i & =  \sum_j \left[ G_{ij} V_j +  \frac{\partial I_i}{\partial V_{Sj}} V_{Sj}\right] + \frac{1}{2}\sum_{jk}\frac{\partial^2 I_i}{\partial V_j \partial V_k}V_jV_k +\dots \label{eq1}\\
I_{Si} & =  \sum_j\left[ G_{ij}^S V_{Sj} +   \frac{\partial I_{Si}}{\partial V_{j}}V_{j}\right] + \frac{1}{2}\sum_{jk}\frac{\partial^2 I_{Si}}{\partial V_{Sj} \partial V_{Sk}}V_{Sj}V_{Sk}+\dots,\label{eq2}
\end{align}
where the derivatives are taken at $\mathbf V=\mathbf V_S= 0$. The coefficients $G_{ij}$ and $G^S_{ij}$ are the usual multi-terminal conductances and spin conductances. The coefficient $\frac{\partial I_i}{\partial V_{Sj}}$ is the linear response of the electric current $I_i$ with respect to a small change in $V_{Sj}$, and similarly for $\frac{\partial I_{Si}}{\partial V_{j}}$. The higher-order derivatives are nonlinear response coefficients.

We are also interested in various cumulants (correlation functions) of the currents. In particular, the noise of the electric current is defined as
\begin{align}
\label{15}
S_{ij} & =2\int dt dt' \overline{(I_i(t)-\bar I_i)(I_j(t')-\bar I_j)}/\tau  \nonumber\\ & =2\int dt dt' [\overline{I_i(t) I_j(t')}-\bar I_i\bar I_j]/\tau,
\end{align}
where the bar means the average with respect to quantum and thermal fluctuations, and the factor of 2 follows a convention. The noise $S^S$ of the spin current and the cross-noise $S^{\rm cr}$ between the spin and electric currents are
\begin{align}
S^{S}_{ij} & =2\int dt dt' [\overline{I_{Si}(t) I_{Sj}(t')}-\bar I_{Si}\bar I_{Sj}]/\tau, \\
S^{\rm cr}_{ij} & =2\int dt dt' [\overline{I_{i}(t) I_{Sj}(t')}-\bar I_{i}\bar I_{Sj}]/\tau.
\end{align}
The third cumulant of the electric current
\begin{align}
\label{16}
C_{ijk}  = & \frac{2}{\tau}\int dt dt' dt''\overline{(I_i(t)-\bar I_i)(I_j(t')-\bar I_j)(I_k(t'')-\bar I_k)} \nonumber\\
= &\frac{2}{\tau}\int dt dt' dt'' \Big(\overline{I_i I_j I_k}-[\overline{I_i I_j}\bar I_k+\overline{I_j I_k}\bar I_i+\overline{I_k I_i}\bar I_j] & & \nonumber\\
 & +2\bar I_i \bar I_j \bar I_k\Big).
\end{align}
We would like to emphasize that the third cumulant is accessible experimentally \cite{3moment}.

It is convenient to rewrite the above definitions in terms of $\Delta Q_i$ and $\Delta S_{zi}$. The cross-noise between $I_i$ and $I_j$ can be written as
\begin{equation}
S_{ij} =  \lim_{\tau\rightarrow \infty}\frac{2}{\tau} \left\langle(\Delta Q_i - \langle{\Delta Q_i}\rangle)(\Delta Q_j - \langle{\Delta Q_j}\rangle)\right\rangle.
\end{equation}
Of particular interest are lower-order coefficients in the Taylor expansion of $S_{ij}$ in the powers of $\mathbf V$ and $\mathbf V_S$. Keeping up to the linear terms in ${\bf V}$ and ${\bf V}_S$, we can approximate $S_{ij}$ by
\begin{align}
S_{ij} &=\lim_{\tau\rightarrow\infty}\frac{2}{\tau}\langle\Delta Q_i\Delta Q_j\rangle \label{noise1}
\end{align}
To obtain (\ref{noise1}) one uses the fact that the average currents vanish at zero bias, i.e., $\langle{\Delta Q_i}\rangle/\tau$ is linear in ${\bf V}, {\bf V}_S$ in the limit of small ${\bf V}, {\bf V}_S$ and large $\tau$.

Similarly, in the linear order in ${\bf V}, {\bf V}_S$, the noises of the spin currents $I_{Si}$ and $I_{Sj}$ and the cross-noises of $I_i$ and $I_{Sj}$ can be expressed as
\begin{align}
S^S_{ij}&=\lim_{\tau\rightarrow\infty}\frac{2}{\tau}\langle\Delta S_{zi}\Delta S_{zj}\rangle, \label{noise2}\\
S^{\rm cr}_{ij}&=\lim_{\tau\rightarrow\infty}\frac{2}{\tau}\langle\Delta Q_i\Delta S_{zj}\rangle. \label{noise3}
\end{align}
In the zeroth order in the bias, the third cumulants of the spin and electric currents can be written as
\begin{align}
C_{ijk}&=-\lim_{\tau\rightarrow\infty}\frac{2}{\tau}\langle \Delta Q_i\Delta Q_j\Delta Q_k\rangle, \label{third1} \\
C_{ijk}^S&=-\lim_{\tau\rightarrow\infty}\frac{2}{\tau}\langle \Delta S_{zi}\Delta S_{zj}\Delta S_{zk}\rangle. \label{third4}
\end{align}

In the bulk of this paper we will focus on the relations between the currents Eq.(\ref{13},\ref{14}), noises (\ref{noise1}-\ref{noise3}) in the linear order in the bias, and the third cumulants  (\ref{third1}, \ref{third4}) at zero bias. One can also define higher-order cumulants and keep their dependence on $\mathbf V, \mathbf V_S$ to an arbitrary order. However, they are experimentally less relevant. We discuss higher-order cumulants in Appendix \ref{sec:ho} for completeness.



\subsection{Derivation of fluctuation relations}

We are now ready to use the fluctuation theorem (\ref{11}) to establish several fluctuation relations for the transport coefficients. These relations include the famous Nyquist formula and the Onsager reciprocal relations for electric currents, and their generalizations to spin currents.

\begin{widetext}

To begin, we introduce the following piece of notation:
\begin{eqnarray}
\label{12}
\mathcal C^{\bf n,\bf m} ({\bf B}, {\bf V}, {\bf V}_S)
 = \lim_{\tau\rightarrow \infty} \frac{1}{\tau}
\langle \prod_i\Delta Q_i^{n_i} \Delta S_{zi}^{m_i}\rangle_{{\bf B},{\bf V},{\bf V}_S}=\lim_{\tau\rightarrow \infty} \frac{1}{\tau}
\sum_{\Delta{\bf Q},\Delta{\bf S}_z}\prod_i\Delta Q_i^{n_i}\Delta S_{zi}^{m_i}\ P(\Delta{\bf Q},\Delta{\bf S}_{z}, {\bf V}, {\bf V}_S, {\bf B}).
\end{eqnarray}
This notation allows a compact and simple formulation of the expressions (\ref{13}, \ref{14}, \ref{noise1}-\ref{third4}). Higher-order cumulants are expressed in terms of combinations of the above quantities (\ref{12}) with different $\mathbf n, \mathbf m$.

We next consider the Taylor expansion of 
$\mathcal C^{\bf n,\bf m} ({\bf B}, {\bf V}, {\bf V}_S)$ near $\mathbf V= \mathbf V_S =0$:
\begin{equation}
\label{19}
\mathcal C^{\bf n,\bf m} ({\bf B}, {\bf V}, {\bf V}_S)
=\sum_{{\boldsymbol \nu}, {\boldsymbol \mu}}L^{{\bf n}, {\bf m}}_{{\boldsymbol \nu},{\boldsymbol \mu}}({\bf B})\prod_i \left(\frac{V_i}{T}\right)^{\nu_i}
\left(\frac{V_{Si}}{T}\right)^{\mu_i}.
\end{equation}
The coefficients $L_{\boldsymbol \nu, \boldsymbol\mu}^{\mathbf n, \mathbf m}(\mathbf B)$ are the transport coefficients that we are interested in. For example, the conductance $G_{ij}$ equals $-T L_{\boldsymbol \nu, \boldsymbol\mu}^{\mathbf n, \mathbf m}(\mathbf B)$ with $n_i=\nu_j=1$ and all other entries of $\mathbf n, \mathbf m, \boldsymbol \nu, \boldsymbol \mu$ vanishing. The spin conductance $G_{ij}^S$ equals $-T L_{\boldsymbol \nu, \boldsymbol\mu}^{\mathbf n, \mathbf m}(\mathbf B)$ with $m_i=\mu_j=1$ and all other entries vanishing. The physical meaning of the other coefficients $L^{{\bf n}, {\bf m}}$ is seen from their relation with various noise cumulants. For example,
$L^{{\bf 0},{{\bf m}_{ijk}}}_{{\bf 0},{\bf 0}}=-C_{ijk}^S/2$, Eq. (\ref{third4}), where the three nonzero entries in ${\bf m}_{ijk}$ are $m_i=m_j=m_k=1$.


Next, we substitute the fluctuation theorem (\ref{11}) in the right hand side of Eq. (\ref{12}) and Taylor expand the right hand side in powers of $ \frac{V_i}{T}$ and $\frac{V_{Si}}{T}$. A comparison of this Taylor series with the expansion (\ref{19}) yields a family of general relations among the transport coefficients at the opposite orientations of the magnetic field
\begin{equation}
\label{20}
L^{{\bf n}, {\bf m}}_{{\boldsymbol \nu},{\boldsymbol \mu}}({\bf B})=(-1)^{\sum_i (n_i+\mu_i)}\sum_{{\bf u}={\bf 0}}^{\boldsymbol \nu}
\sum_{{\bf w}={\bf 0}}^{\boldsymbol \mu}\prod_i\left(\frac{1}{u_i!w_i!}\right)L^{{\bf n+u}, {\bf m+w}}_{{\boldsymbol \nu}-\mathbf u,
{\boldsymbol \mu}-\mathbf w}(-{\bf B}),
\end{equation}
where ${\boldsymbol \nu}-\boldsymbol u$ denotes the vector with the components $\nu_i-u_i$, and the summation $\sum_{{\bf u}={\bf 0}}^{\boldsymbol \nu}=\sum_{u_1=0}^{\nu_1}\sum_{u_2=0}^{\nu_2}\cdots$.
In the following sections, we will use Eq. (\ref{20}) to extract fluctuation relations for experimental observables in various settings.



\end{widetext}


\section{Examples}
\label{sec:example}

We now apply the fluctuation relations (\ref{20}) to several physical systems.
Our focus is on relations between the experimentally most relevant transport quantities, including currents, noises and the third cumulants.
In this section we assume that all reservoir temperatures are the same. The spin Seebeck effect is addressed in the next section.

\subsection{Two-terminal setup}
\label{sec:two_terminal}

\begin{figure}[b]
\includegraphics[width=3.3in]{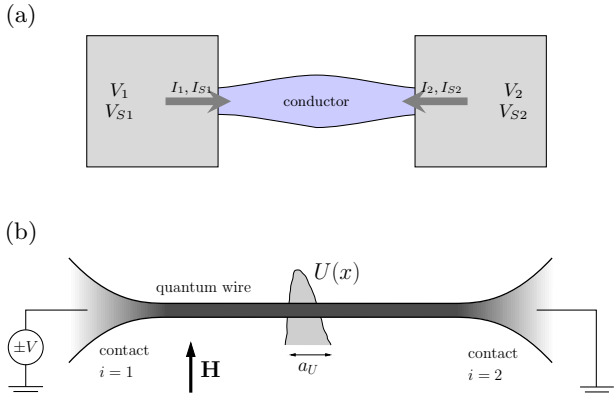}
\caption{(a) Two-terminal setup for charge and spin transport. The arrows represent our convention about the positive directions of the currents. (b) A spin-current rectifier [from Ref.~\onlinecite{rec6}].}\label{fig_twoterminal}
\end{figure}


Let us consider a two-terminal conductor, i.e., a system with two reservoirs [Fig. \ref{fig_twoterminal}(a)]. This setting is relevant for spin-current rectification effect that emerges due to nonlinear spin transport in a conductor.
Fig.~\ref{fig_twoterminal}(b) illustrates the spin-current rectifier from Ref. \onlinecite{rec6}. It includes a quantum wire with an asymmetric potential barrier in a uniform magnetic field.

First of all, we notice that charge conservation implies $\Delta Q_1 + \Delta Q_2=0$. This leads to $I_1+I_2=0$. Therefore, we have three independent currents, $I_1, I_{S1}, I_{S2}$. (Recall that spin may not be conserved in the conductor. $I_{S1}$ and $I_{S_2}$ are defined at the interfaces between the conductor and terminals.) Besides, gauge invariance tells us that transport quantities depend only on the voltage difference $V_1-V_2$ between the reservoirs and not on the absolute values of their chemical potentials. Thus, without loss of generality, we are allowed to set $V_2=0$. Therefore, we have three independent biases $V_1, V_{S1}, V_{S2}$ to drive the currents.

Consider linear transport for a warm-up exercise. From the general fluctuation relation (\ref{20}), we derive in Appendix \ref{app1} the following relations for the noises
\begin{align}
S_{11}(\mathbf B) & = 4T\frac{\partial I_1(\mathbf B)}{\partial V_{1} }, \label{fdt1} \\
S_{11}^S(\mathbf B)&  = 4T\frac{\partial I_{S1}(\mathbf B)}{\partial V_{S1} },\label{fdt2} \\
S_{11}^{\rm cr}(\mathbf B) &  = 2T\left[\frac{\partial I_1(\mathbf B)}{\partial V_{S1} } +\frac{\partial I_{S1}(\mathbf B)}{\partial V_1}\right],\label{fdt3}
\end{align}
and the linear conductances
\begin{align}
\frac{\partial I_{1}(\mathbf B)}{\partial V_{1}} & = \frac{\partial I_{1}(-\mathbf B)}{\partial V_{1}},\label{or1}\\
\frac{\partial I_{S1}(\mathbf B)}{\partial V_{S1}} & = \frac{\partial I_{S1}(-\mathbf B)}{\partial V_{S1}},\label{or2}\\
\frac{\partial I_{1}(\mathbf B)}{\partial V_{S1}} & = -\frac{\partial I_{S1}(-\mathbf B)}{\partial V_{1}}, \label{or3}
\end{align}
where all quantities are evaluated at equilibrium $\mathbf V=\mathbf V_S =0$. The relation (\ref{fdt1}) is the famous Nyquist formula, and (\ref{fdt2}-\ref{fdt3}) are its generalizations for spin currents. The relation (\ref{or1}) is the Onsager reciprocity relation, and (\ref{or2}-\ref{or3}) are its generalizations for spin currents. Of particular interest is the relation (\ref{or3}). It has an additional minus sign compared to (\ref{or1}) and (\ref{or2}). This additional sign originates from the fact that spin is odd under time reversal.


Let us now turn to the nonlinear transport. We show in Appendix \ref{app1} that
\begin{align}
C_{111}(\mathbf B) & =3T\frac{\partial S_{11}(\mathbf B)}{\partial V_1} - 6 T^2 \frac{\partial^2 I_1(\mathbf B)}{\partial V_1^2},  \label{t1}\\
C_{111}^S(\mathbf B) & = 3T \frac{\partial S_{11}^S(\mathbf B)}{\partial V_{S1}} -6T^2 \frac{\partial^2 I_{S1}(\mathbf B)}{\partial V_{S1}^2}, \label{t2}\\
C_{111}(\mathbf B) &  = T \frac{\partial S_{11}^-}{\partial V_1}\label{t3},\\
C_{111}^S (\mathbf B) & = T \frac{\partial S_{111}^{S,+}}{\partial V_{S1}} \label{t4}
\end{align}
and
\begin{align}
\frac{\partial S_{11}^+}{\partial V_1} & = 2T\frac{\partial^2 I_1^+ }{\partial V_1^2}, \quad
\frac{\partial S_{11}^-}{\partial V_1}  = 6T\frac{\partial^2 I_1^- }{\partial V_1^2},\label{nc1} \\
\frac{\partial S_{11}^{S,+}}{\partial V_{S1}}  &= 6T\frac{\partial^2 I_{S1}^{+} }{\partial V_{S1}^2},\quad
\frac{\partial S_{11}^{S,-}}{\partial V_{S1}}  = 2T\frac{\partial^2 I_{S1}^{-} }{\partial V_{S1}^2}, \label{nc2}
\end{align}
where we use the notation $\mathcal O^\pm = \mathcal O(\mathbf B) \pm \mathcal O(-\mathbf B)$ for convenience. Furthermore, we show in Appendix \ref{app1} that
\begin{align}
-2T\frac{\partial^2 I_{1}(\mathbf B)}{\partial V_{1}^2} &= \frac{\partial S_{11}(-\mathbf B)}{\partial V_{1}}-4T\frac{\partial^2 I_{1}(-\mathbf B)}{\partial V_1^2}, \label{crosstwo1}\\
2T\frac{\partial^2 I_{S1}(\mathbf B)}{\partial V_{S1}^2}& =\frac{\partial S^S_{11}(-\mathbf B)}{\partial V_{S1}}-4T \frac{\partial^2 I_{S1}(-\mathbf B)}{\partial V_{S1}^2}, \label{crosstwo3}
\end{align}
and
\begin{align}
-2T\frac{\partial^2 I_{1}(\mathbf B)}{\partial V_{S1}^2} &=\frac{\partial S^S_{11}(-\mathbf B)}{\partial V_1}-4T\frac{\partial^2 I_{S1}(-\mathbf B)}{\partial V_{1}\partial V_{S1}},\label{crosstwo2} \\
2T\frac{\partial^2 I_{S1}(\mathbf B)}{\partial V_{1}^2} &= \frac{\partial S_{11}(-\mathbf B)}{\partial V_{S1}}-4T\frac{\partial^2 I_{1}(-\mathbf B)}{\partial V_{S1}\partial V_{1}}. \label{crosstwo4}
\end{align}
Again, all quantities are evaluated at $\mathbf V=\mathbf V_S =0$. One can derive an infinite number of additional fluctuation relations. We focus on the above results because of their simplicity.
In particular, only currents and second noise cumulants enter Eqs. (\ref{nc1}-\ref{crosstwo4})

The relations (\ref{t1}),(\ref{t3}) and (\ref{nc1}) for the third cumulant and nonlinear transport coefficients have been derived before\cite{saito}, and the relations (\ref{t2}), (\ref{t4}) and (\ref{nc2}) are their generalizations for spin currents.  Note that according to (\ref{t3}), $C_{111}(\mathbf B)$ is odd under $\mathbf B \rightarrow - \mathbf B$, while $C_{111}^S (\mathbf B)$ is even. We notice that the right hand sides of Eqs.~(\ref{crosstwo1}-\ref{crosstwo4}) can be obtained by differentiating the differences of  the left and right hand sides of the Nyquist formulas (\ref{fdt1},\ref{fdt2}). Therefore, Eqs. (\ref{crosstwo1}-\ref{crosstwo4}) can be understood as relations between the nonlinear conductances and the degree of the violation of the Nyquist formulas away from thermal equilibrium. Finally, it is worth to mention that the odd parity of spin under time reversal leads to various differences
(not just different signs on the left!) between (\ref{t3}) and (\ref{t4}), (\ref{nc1}) and (\ref{nc2}), (\ref{crosstwo1}) and (\ref{crosstwo3}),  and (\ref{crosstwo2}) and (\ref{crosstwo4}).



\subsection{Multi-terminal setup}
\label{sec:multi_terminal}

\begin{figure}[b]
\includegraphics{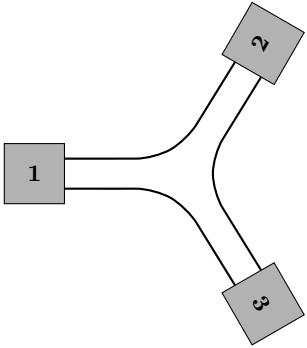}
\caption{Sketch of a three-terminal setup.}\label{fig_threeterminal}
\end{figure}

The fluctuation relations for two-terminal setups presented in the previous subsection can be generalized to multi-terminal setups. Below, we list the relations analogous to the above two-terminal relations (see Appendix \ref{app2} for derivations). The three-terminal geometry, illustrated in Fig.~\ref{fig_threeterminal}, is particularly interesting due to its relevance to transistors \cite{sptr1,sptr2}.

First of all, the generalizations of the Nyquist formula read
\begin{align}
S_{ij}(\mathbf B) &= 2T \left[\frac{\partial I_i(\mathbf B)}{\partial V_j} +\frac{\partial I_j(\mathbf B)}{\partial V_i} \right], \label{multi1}\\
S_{ij}^S(\mathbf B)& = 2T \left[\frac{\partial I_{Si}(\mathbf B)}{\partial V_j} +\frac{\partial I_{Sj}(\mathbf B)}{\partial V_i} \right], \\
S_{ij}^{\rm cr}(\mathbf B) & = 2T \left[\frac{\partial I_i(\mathbf B)}{\partial V_{Sj}} +\frac{\partial I_{Sj}(\mathbf B)}{\partial V_i} \right]. \label{multifdt}
\end{align}
The Onsager reciprocity relations read
\begin{align}
\frac{\partial I_i(\mathbf B )}{\partial V_j} & = \frac{\partial I_j(-\mathbf B )}{\partial V_i}, \\
\frac{\partial I_{Si}(\mathbf B )}{\partial V_{Sj} } & = \frac{\partial I_{Sj}(-\mathbf B )}{\partial V_{Si}} ,\label{multior2} \\
\frac{\partial I_i(\mathbf B )}{\partial V_{Sj}} & = -\frac{\partial I_{Sj}(-\mathbf B )}{\partial V_i}. \label{multior}
\end{align}
We emphasize the minus sign on the right hand side of (\ref{multior}). Relations, analogous to (\ref{t1},\ref{t2}), read
\begin{align}
C_{ijk}(\mathbf B) = & T\frac{\partial S_{ij}(\mathbf B)}{\partial V_k}  - 2 T^2 \frac{\partial^2 I_i(\mathbf B)}{\partial V_j\partial V_k}+ \ c.p., \label{m0}\\
C_{ijk}^S(\mathbf B) = &T\frac{\partial S_{ij}^S(\mathbf B)}{\partial V_{Sk}} - 2 T^2\frac{\partial^2 I_{Si}(\mathbf B)}{\partial V_{Sj}\partial V_{Sk}}+ \ c.p.,\label{m1}
\end{align}
where ``$c.p.$'' stands for the cyclic permutations of the indices $i,j,k$. Relations analogous to (\ref{t3}) and (\ref{t4}) read
\begin{align}
C_{ijk}(\mathbf B) &  = T \frac{\partial S_{ij}^-}{\partial V_k}=T \frac{\partial S_{jk}^-}{\partial V_i}=T \frac{\partial S_{ki}^-}{\partial V_j},\label{m5}\\
C_{ijk}^S (\mathbf B) & = T \frac{\partial S_{ij}^{S,+}}{\partial V_{Sk}}=T \frac{\partial S_{jk}^{S,+}}{\partial V_{Si}}=T \frac{\partial S_{ki}^{S,+}}{\partial V_{Sj}},\label{m2}
\end{align}
and relations analogous to (\ref{nc1}) and (\ref{nc2}) read
\begin{align}
\frac{\partial S_{ij}^+}{\partial V_{k}} + c. p. & = 2T\left(\frac{\partial^2 I_i^+ }{\partial V_j\partial V_{k}} +c.p.\right), \\
\frac{\partial S_{ij}^-}{\partial V_{k}} + c. p. & = 6T\left(\frac{\partial^2 I_i^- }{\partial V_j\partial V_{k}} +c.p.\right),\\
\frac{\partial S_{ij}^{S,+}}{\partial V_{Sk}} + c. p. & = 6T\left(\frac{\partial^2 I_{Si}^+ }{\partial V_{Sj}\partial V_{Sk}} +c.p.\right),\label{m3}\\
\frac{\partial S_{ij}^{S,-}}{\partial V_{Sk}} + c. p. & = 2T\left(\frac{\partial^2 I_{Si}^- }{\partial V_{Sj}\partial V_{Sk}} +c.p.\right).\label{m4}
\end{align}
The relations analogous to (\ref{crosstwo1}, \ref{crosstwo3}) read
\begin{align}
\frac{\partial^2 I_{k}(\mathbf B)}{\partial V_{i}\partial V_{j}} &=-\frac{1}{2T} \frac{\partial S_{ij}(-\mathbf B)}{\partial V_{k}}+\frac{\partial^2 I_{i}(-\mathbf B)}{\partial V_k\partial V_{j}} + \frac{\partial^2 I_{j}(-\mathbf B)}{\partial V_k\partial V_{i}},  \label{cross1}\\
\frac{\partial^2 I_{Sk}(\mathbf B)}{\partial V_{Si}\partial V_{Sj}}& =\frac{1}{2T} \frac{\partial S^S_{ij}(-\mathbf B)}{\partial V_{Sk}}-\frac{\partial^2 I_{Si}(-\mathbf B)}{\partial V_{Sk}\partial V_{Sj}} - \frac{\partial^2 I_{Sj}(-\mathbf B)}{\partial V_{Sk}\partial V_{Si}}  \label{cross3}
\end{align}
and the relations analogous to (\ref{crosstwo2}, \ref{crosstwo4}) read
\begin{align}
\frac{\partial^2 I_{k}(\mathbf B)}{\partial V_{Si}\partial V_{Sj}} &=-\frac{1}{2T} \frac{\partial S^S_{ij}(-\mathbf B)}{\partial V_k}+\frac{\partial^2 I_{Si}(-\mathbf B)}{\partial V_k\partial V_{Sj}} + \frac{\partial^2 I_{Sj}(-\mathbf B)}{\partial V_k\partial V_{Si}},\label{cross2} \\
\frac{\partial^2 I_{Sk}(\mathbf B)}{\partial V_{i}\partial V_{j}} &=\frac{1}{2T} \frac{\partial S_{ij}(-\mathbf B)}{\partial V_{Sk}}-\frac{\partial^2 I_{i}(-\mathbf B)}{\partial V_{Sk}\partial V_{j}} - \frac{\partial^2 I_{j}(-\mathbf B)}{\partial V_{Sk}\partial V_{i}}.  \label{cross4}
\end{align}
All the quantities above are evaluated at $\mathbf V = \mathbf V_S = 0$. The fluctuation relations hold for any indices $i,j,k=1,\dots,N$, where $N$ is the number of the terminals ($N=3$ in Fig. \ref{fig_threeterminal}).


\subsection{Currents of spin-up and -down electrons}
We have focused above on the spin and charge currents and noises. It is also of interest to consider the currents and noises of spin-up and -down electrons. The latter currents are just linear combinations of electric and spin currents. We derive some fluctuation relations in this language which facilitates the comparison of our results to those of Ref.~\onlinecite{lopez12} in the next subsection.

We will denote the currents of spin-up and -down electrons as $I_{i\alpha}$, where $\alpha=+$ corresponds to the spin-up current and $\alpha=-$ corresponds to the spin-down current. The charges of the spin-up and -down electrons in reservoir $i$ are $Q_{i\alpha}=Q_i/2-\alpha eS_{zi}$. The conjugate chemical potentials $V_{i\alpha}=V_i-\frac{\alpha}{2e}V_{Si}$. Under time-reversal, $Q_{i\alpha}\rightarrow Q_{i\bar\alpha}$, $V_{i\alpha}\rightarrow V_{i\bar\alpha}$, where $\overline +=-$, $\overline -=+$. The fluctuation theorem (\ref{11}) becomes
\begin{equation}
\label{28}
P(\Delta {\bf Q}, {\bf V}, {\bf B})=\exp\left[-\frac{{\bf V}\cdot{\Delta{\bf Q}}}{T}\right ]P(-\Delta {\overline{\bf Q}}, {\overline{\bf V}}, -{\bf B}),
\end{equation}
where $\bf Q$ and $\bf V$  stay for the vectors with the components $Q_{i\alpha}$ and $V_{i\alpha}$, $\overline{\bf Q}$ and $\overline {\bf V}$ are the vectors with the components $Q_{i\bar\alpha}$ and $V_{i\bar\alpha}$, and ${\bf V}\cdot{\Delta{\bf Q}}=\sum_{i\alpha} V_{i\alpha}\Delta Q_{i\alpha}$.

Below we derive two relations that connect the nonlinear conductance with the noise  and the third cumulant. We sum the right and left hand sides of Eq. (\ref{28}) over all possible $\Delta\bf Q$. We get
$\langle 1\rangle_{{\bf V},{\bf B}}=1$ on the left and expand the sum on the right to the third order in ${\bf V}$. We find
\begin{eqnarray}
\label{29}
C_{i\alpha,j\beta,k\gamma}({\bf V}= {\bf 0},{\bf B})=
T\left[\frac{\partial S_{j\beta,k\gamma}({\bf V}={\bf 0},{\bf B})}{\partial V_{i\alpha}}+{c.p.}\right] & &\nonumber\\
-2T^2\left[\frac{\partial^2 I_{k\gamma}({\bf V}={\bf 0},{\bf B})}{\partial V_{i\alpha}\partial V_{j\beta}} +c.p.\right], & &
\end{eqnarray}
where $c.p.$ stays for the cyclic permutations of the pairs of the indices $i\alpha$, $j\beta$ and $k\gamma$.


We next use Eq. (\ref{28}) to compute $\langle\Delta Q_{i\bar\alpha}\rangle_{{\bf V},{\bf B}}-\langle\Delta Q_{i\alpha} \rangle_{\overline{\bf V}, -{\bf B}}$. We then subtract the right hand side of the equation from its left hand side and expand to the second order in ${\bf V}$. The result is
\begin{eqnarray}
\label{30}
C _{i\bar\alpha,j\bar\beta,k\bar\gamma}({\bf V}={\bf 0}, {\bf B})-C _{i\alpha,j\beta,k\gamma}({\bf V}={\bf 0}, -{\bf B})= & & \nonumber\\
T\Big[\frac{\partial S_{i\bar\alpha,j\bar\beta}({\bf V}={\bf 0}, {\bf B})}{\partial V_{k\bar\gamma}}+\frac{\partial S_{i\bar\alpha,k\bar\gamma}({\bf V}={\bf 0}, {\bf B})}{\partial V_{j\bar\beta}}
& & \nonumber\\
-\frac{\partial S_{i\alpha,j\beta}({\bf V}={\bf 0}, -{\bf B})}{\partial V_{k\bar\gamma}}-\frac{\partial S_{i\alpha,k\gamma}({\bf V}={\bf 0}, -{\bf B})}{\partial V_{j\bar\beta}}\Big]. & &
\end{eqnarray}


\subsection{Comparison with existing results}
\label{sec:comparison}

In this section, we compare our result with the existing work, Refs.~\onlinecite{spin-utsumi} and \onlinecite{lopez12}, on fluctuation relations for spin currents and other multi-component currents.

Ref.~\onlinecite{spin-utsumi} investigates a multi-component current of time-reversal-invariant quantities. At the same time, we employ the fact that spin is odd under time reversal.
This fact manifests itself in the sign factor $(-1)^{\sum_i (n_i+\mu_i)}$ in the general expression ~(\ref{20}) and in the sign factors in the fluctuation relations (\ref{or3}), (\ref{multior}), etc. If spin were even the sign factor in (\ref{20}) would be $(-1)^{\sum_i (n_i+m_i)}$. Note that Ref.~\onlinecite{spin-utsumi} derives a similar relation to (\ref{eq54}) but with no minus sign [the minus sign in (\ref{eq54}) is carried over to (\ref{or3})]. This illustrates the difference of the fluctuation relations for spin currents from  the fluctuation relations for multi-components currents of time-reversal-invariant quantities.

In Ref.~\onlinecite{lopez12}, the authors derived several fluctuation relations for spin current without using microreversibility. In contrast, our derivation depends heavily on microreversibility. For a detailed discussion of microreversibility, see Appendix \ref{sec:microreversibility}.

Our relation (\ref{29}) is similar to the fluctuation relations from Refs. \onlinecite{forster08,lopez12} and could be obtained without the use of microreversibility. On the other hand, Eq. (\ref{30}) is new and could not be obtained with the methods of Refs. \onlinecite{forster08,lopez12}. Indeed, summing Eq. (\ref{30}) over all possible choices of the spin indices and taking the limit ${\bf V}=\overline{\bf V}$, one reproduces a fluctuation relation in Ref.\onlinecite{saito} for electric currents. The latter relation cannot be obtained without microreversibility and the same is true for our result (\ref{30}).

\section{Thermospin transport}
\label{sec:thermo}
We have assumed identical temperatures in the reservoirs in the above discussion. In this section, we consider the case of the reservoirs at different temperatures. The temperature gradient leads
to the spin Seebeck effect \cite{see1,see2,see3}. In the ordinary Seebeck effect, an electric current flows between two conductors at the same voltage but different temperatures. In the absence of the symmetry between the two projections of spin, the currents of spin-up and -down electrons in response to a thermal gradient are not the same. Thus, a spin current is generated.

Let us start with extending the fluctuation theorem (\ref{11}) to the case of the reservoirs with different temperatures. Consider the same setup as in Sec.~\ref{ft}, a conductor attached to several large reservoirs. The reservoirs are maintained at different temperatures $T_i$, different electro-chemical potentials $V_i$, and different spin chemical potentials $V_{Si}$. We follow the same measurement protocol as before: we measure the total energy, total charge and the $z$-component of the total spin of each reservoir before and after an evolution period $\tau$. Following a similar derivation to the one in Sec.~\ref{ft}, one can show that
\begin{align}
 & \frac{P(\Delta{\bf E}, \Delta{\bf Q}, {\Delta{\bf S}_{z}}, {\bf T}, {\bf V}, {\bf V}_S, {\bf B})}{P(-\Delta{\bf E},-\Delta{\bf Q}, \Delta{\bf S}_z, {\bf T}, {\bf V}, -{\bf V}_S, -{\bf B})} \nonumber \\
& \quad\quad =  \exp\left[\sum_i \left(\beta_i\Delta E_i +\xi_i \Delta Q_i + \zeta_i\Delta S_{zi}\right )\right], \label{thermo1}
\end{align}
where $P$ is the probability distribution function for observing the change $\Delta E_i$ of the energy, the change  $\Delta Q_i$ of the charge, and the change $\Delta S_{zi}$ of the $z$-component of the spin  in reservoir $i$ during the evolution period $\tau$, with given $T_i, V_i, V_{Si}$ and the magnetic field $\mathbf B$. The quantities $\beta_i, \xi_i, \zeta_i$ are defined as $\beta_i = 1/T_i$, $\xi_i = - V_i/T_i$ and $\zeta_i = - V_{S_i}/T_i$.

We are now in the position to derive fluctuation relations for thermospin transport. One can derive as many relations as in Sec.~\ref{sec:example}. We will focus only on the relations that crucially depend on the odd parity of spin. As a warming up exercise, we consider the Onsager relations in linear transport. A detailed discussion of the Onsager relations in the spin Seebeck effect can be found in Ref. \onlinecite{see2}. The approach of Ref. \onlinecite{see2} is very different from ours and builds on the Landauer-B\"uttiker formalism.

The average heat current, injected into the conductor from reservoir $i$, is given by
\begin{equation}
I_{hi} =-\lim_{\tau \rightarrow \infty}\frac{1}{\tau}\langle \left(\Delta E_i - V_i \Delta Q_i - V_{Si} \Delta S_{zi}\right) \rangle_{\mathbf B,\mathbf T, \mathbf V, \mathbf V_S}. \label{thermo3}
\end{equation}
This definition reflects the fact that only the excess energy change $\Delta E_i- V_i \Delta Q_i - V_{Si} \Delta S_{zi}$ is dissipated into heat; $V_i \Delta Q_i + V_{Si} \Delta S_{zi}$ can be thought of as the potential energy.
We also find it convenient to define the energy currents
\begin{equation}
I_{Ei} = -\lim_{\tau \rightarrow \infty}\frac{1}{\tau}\langle  \Delta E_i  \rangle_{\mathbf B,\mathbf T, \mathbf V, \mathbf V_S}.
\end{equation}
Let us now consider Eq. (\ref{thermo1}) in the limit of zero voltage bias, $\xi_i=0$. Eq. (\ref{thermo1}) has now the same structure as Eq. (\ref{11}) with $\beta_i$ in place of $-V_i/T$ and $\zeta_i$ in place of $-V_{Si}/T$.
It follows from the analogy with Eq. (\ref{or3}) that
\begin{equation}
\label{df2}
\frac{\partial I_{E_i}(\mathbf B)}{\partial \zeta_j} = -\frac{\partial I_{Sj}(-\mathbf B)}{\partial \beta_i},
\end{equation}
where the derivatives are taken at equilibrium: $V_{Si}=0$ and $T_i=T$. Making use of the connection between $\beta_i,\zeta_i$ and $T_i, V_{Si}$, the definition of $I_{hi}$ and the fact that spin currents vanish in equilibrium, we obtain the Onsager relation
\begin{align}
\frac{\partial I_{hi}(\mathbf B)}{\partial V_{Sj}}  =  - T \frac{\partial I_{Sj}(-\mathbf B)}{\partial T_i}. \label{thermo2}
\end{align}
Note that the right hand side describes the linear spin Seebeck effect, while the left hand side describes the linear {\it inverse} spin Seebeck effect.

We now turn to the nonlinear Seebeck effect and thus go beyond Ref. \onlinecite{see2}. Below we establish relations analogous to (\ref{cross2}) and (\ref{cross4}). Again, we make use of the similarity between the fluctuation theorems (\ref{11}) and (\ref{thermo1}) in the case of $\xi_i=0$. Taking Eq. (\ref{cross2}) and making the substitutions $-V_i/T\rightarrow\beta_i$, $-V_{Si}/T\rightarrow \zeta_i$ and $I_{i}\rightarrow I_{Ei}$, we find that
\begin{equation}
\frac{\partial^2 I_{Ek}(\mathbf B)}{\partial \zeta_i\partial \zeta_j} =\frac{1}{2} \frac{\partial S^S_{ij}(-\mathbf B)}{\partial \beta_k}+\frac{\partial^2 I_{Si}(-\mathbf B)}{\partial \beta_k\partial \zeta_j} + \frac{\partial^2 I_{Sj}(-\mathbf B)}{\partial \beta_k\partial \zeta_i}.
\end{equation}
Using the definition of $I_{hi}$ and the connection between $\beta_i,\zeta_i$ and $T_i, V_{Si}$, we obtain the relation
\begin{equation}
\frac{1}{T}\frac{\partial^2 I_{hk}(\mathbf B)}{\partial V_{Si}\partial V_{Sj}} =- \frac{1}{2T}\frac{\partial S^S_{ij}(-\mathbf B)}{\partial T_k}+\frac{\partial^2 I_{Si}(-\mathbf B)}{\partial T_k\partial V_{Sj}} + \frac{\partial^2 I_{Sj}(-\mathbf B)}{\partial T_k\partial V_{Si}}. \label{thermo4}
\end{equation}
where the derivatives are computed at $V_i=V_{Si}=0$ and $T_i=T$. During the derivation, several terms are cancelled due to the Onsager relation (\ref{or2}). By similar arguments, it follows from (\ref{cross4}) and
(\ref{multifdt}) that

\begin{align}
T&\frac{\partial^2 I_{Sk}(\mathbf B)}{\partial T_i\partial T_j} +2\delta_{ij}\frac{\partial I_{Sk}(\mathbf B)}{\partial T_j} \nonumber\\
 & = \frac{1}{2T^2} \frac{\partial S_{ij}^h(-\mathbf B)}{\partial V_{Sk}} - \frac{\partial^2 I_{hi}(-\mathbf B)}{\partial V_{Sk}\partial T_j}-  \frac{\partial^2 I_{hj}(-\mathbf B)}{\partial V_{Sk}\partial T_i}\label{thermo5}
\end{align}
where $S^h_{ij}$ is the cross noise between the heat currents $I_{hi}$ and $I_{hj}$. The right hand sides of (\ref{thermo4}) and (\ref{thermo5}) differ by a minus sign, which again manifests the odd parity of spin under time reversal. We note that there is an additional term proportional to $\delta_{ij}$ on the left hand side of (\ref{thermo5}) compared to (\ref{thermo4}).

\section{Summary}
\label{sec:con}

In conclusion, we prove a general fluctuation theorem for spin currents. It imposes a great number of restrictions on transport coefficients and fluctuations of the spin, charge and heat currents in linear and nonlinear transport.  We have focused on the fluctuation relations for currents, noises and third cumulants, and their lower order derivatives. Several relations for the transport, driven by the gradients of chemical potentials, are derived in Sec.~\ref{sec:two_terminal} and Sec.~\ref{sec:multi_terminal}. Sec.~\ref{sec:thermo} addresses thermospin transport.


Our derivation relies on the combination of microreversibility with the assumption that the $z$-component of spin conserves approximately in large parts of the reservoirs near the conductor. If the latter assumption does not hold then it may not be even meaningful to speak about the spin currents, injected from the reservoirs into the conductor. Note that very long spin-coherence times have been reported for donor spins in silicon \cite{silicon}.

Our results do not depend on a particular model and thus apply to many systems. One example is the spin current rectifier \cite{rec6,utsumi15}.  Since our fluctuation relations are exact, they can be employed for testing approximations in theoretical calculations.

The discussion in the preceding sections is most directly connected with mesoscopic conductors that can carry both spin and electric currents. At the same time, the presence of an electric current is not essential for the validity of our results. They also apply to pure spin currents. For example, our fluctuation relations hold for spin diffusion in insulators \cite{spin-diffusion}. The above calculations assume that the spin current is carried
by particles with spin $1/2$. It is straightforward to extend our results to higher spins.
Another interesting setting is topological matter with chiral transport \cite{wang3}. We expect our fluctuation relations to simplify considerably in such systems since chirality implies the vanishing of many transport coefficients \cite{wang2}. A simple example is discussed in Appendix \ref{sec:chiral}.

\begin{acknowledgments}
CW acknowledges the support by the NSF under Grant No. DMR-1254741.
DEF was supported in part by the NSF under Grant No. DMR-1205715.
\end{acknowledgments}

\appendix

\section{Microreversibility}
\label{sec:microreversibility}
\begin{figure}[b]
\includegraphics{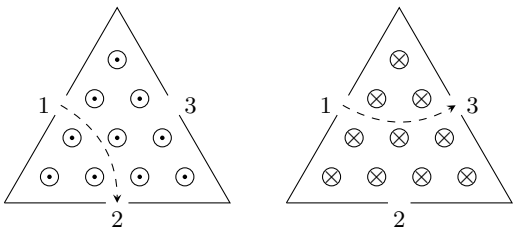}
\caption{Trajectories of particles  in opposite magnetic fields.}\label{fig_mr}
\end{figure}

The principle of microreversibility is expressed by Eq. (\ref{8}). Its physical meaning is simple: after reversing the directions of the magnetic field and the velocities and spins of all particles, the system traces its evolution backwards in time. This is a straightforward consequence of the laws of quantum mechanics. At the same time, the calculations \cite{forster08,forster09}, based on the Landauer-B\"uttiker formalism, seem to conflict with the principle.
In this section we clarify the origin of the conflict. We will see that while the Landauer-B\"uttiker formalism works well in the linear response regime, its application to fluctuations in nonlinear transport faces challenges.

The Landauer-B\"uttiker formalism \cite{datta} uses a single-particle language. Electrons are treated as non-interacting particles in a self-consistent electrostatic potential. The self-consistent potential is different for the opposite directions of the magnetic field \cite{as1,as2}. The origin of that asymmetry is illustrated in Fig.~\ref{fig_mr}. In that example, charged particles enter the conductor through terminal 1. Their trajectories depend on the magnetic field. Depending on its sign, the particles exit through terminal 2 or 3. The average charge density $\rho({\bf r},{\bf B})$ is nonzero only along the particle trajectories. Hence, $\rho({\bf r},{\bf B})\ne\rho({\bf r},-{\bf B})$. Since the self-consistent electrostatic potential $\phi$ is determined by the charge density,
\begin{equation}
\label{df1}
\phi({\bf r},{\bf B})\ne\phi({\bf r},-{\bf B}).
\end{equation}

The central quantity in the Landauer-B\"uttiker formalism is the single-particle scattering matrix \cite{datta}. Refs. \onlinecite{forster08,forster09} compare the scattering amplitude from the initial state with the momentum ${\bf k}_i$ to the final state with the momentum ${\bf k}_f$ for an electron in the magnetic field ${\bf B}$ in the self-consistent potential $\phi({\bf r},{\bf B})$ with the ``time-reversed'' scattering amplitude from the state with the momentum $-{\bf k}_f$ to the momentum $-{\bf k}_i$ in the field $-{\bf B}$ in the potential $\phi({\bf r},-{\bf B})$. If the two self-consistent potentials $\phi({\bf r},{\bf B})$ and $\phi({\bf r},-{\bf B})$ equaled the two scattering amplitudes would be complex conjugate. Eq. (\ref{df1}) implies that they are not in fact complex conjugate. This was interpreted as microreversibility breaking \cite{forster08}. Indeed, the calculations
 \cite{forster08,forster09} with such scattering amplitudes contradict the results from microreversibility \cite{saito}.

The origin of this apparent breakdown can be best understood if we forget about quantum mechanics and consider a classical many-body system. Let us first use a self-consistent approximation. We find a self-consistent electrostatic potential $\phi({\bf r},{\bf B})$ and consider separately the motion of each electron in this potential. Let us now reverse the direction of the final velocity of the electron and the direction of the magnetic field ${\bf B}\rightarrow -{\bf B}$. If the self-consistent potential remained the same then the electron would trace the same trajectory backwards. A different self-consistent potential $\phi({\bf r},-{\bf B})$  implies a different trajectory, i.e., microreversibility breaks down. How does this compare with the exact solution of the equations of motion of a many body system?  The potential energy of a system of electrons depends only on their positions and does not depend on the magnetic field. It follows then easily that after one reverses the magnetic field and the velocities of {\it all} electrons then each electron traces its trajectory backwards, exactly as the microreversibility principle demands. In other words, microreversibility breaking is an artifact of the single-particle self-consistent approximation inherent to the Landauer-B\"uttiker formalism. Microscopic calculations for many-body quantum-mechanical models beyond the Landauer-B\"uttiker approximation \cite{saito09,nasb10} do agree with the predictions \cite{saito} from microreversibility for charge transport. The goal of this work is to investigate the consequences of microreversibility for nonlinear spin transport.

The above discussion shows that a magentic field does not destroy microrevesibilty if it is present without a magnetic field. Microrevesibility without a magnetic field has been addressed by many authors\cite{Cft, n1, n2, n3}. The results, based on microreversibility, can be expected to apply in a Hamiltonian system, where the conductor can only exchange energy with the reservoirs. This condition cannot be relaxed if the reservoirs are maintained at different temperatures. When all temperatures are the same such energy conservation is no longer necessary. Indeed, the insulating enviroment can be included into one of the reservoirs in all derivations. Another condition remains crucial for the validity of our results even when all temperatures equal as discussed in Refs.~\onlinecite{wang3, n4}: the charge transfer must be small in the detector that probes the system. Otherwise the fluctuation relation is modifed \cite{n4}.

\section{Detection of spin currents $I_S$ and chemical potentials $V_S$.}

\label{sec:spinchemopoten}
\begin{figure}
\includegraphics{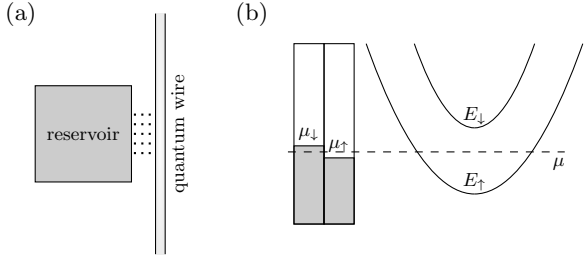}
\caption{Detecting spin chemical potentials. (a) A quantum wire proximate to the reservoir as a probe. (b) Schematics of the energy subbands of the wire and the relative values of different energies.}\label{fig_wire}
\end{figure}

In order to connect our results with spintronic experiments, we briefly review the ways to measure the spin currents $I_{Si}$ and spin chemical potentials $V_{Si}$ in this appendix.

Detecting spin currents is of crucial importance for spintronics and numerous methods to do so where implemented and/or proposed. Examples include the use of the spin-current induced Hall effect \cite{IS1}, magnetic resonance
\cite{IS2} and optical techniques \cite{IS3}. The measurement of the chemical potentials $V_S$ reduces to the measurement of electric and spin currents. Below we discuss one way to extract $V_S$ from currents.

The setup is illustrated in Fig.~\ref{fig_wire}. A quantum wire is brought close to reservoir $i$ so that electrons can tunnel between the wire and reservoir.
As discussed below, the tunneling current contains information about the chemical potentials $\mu_\uparrow=-eV_{\uparrow}$ and $\mu_\downarrow=-eV_{\downarrow}$ of the spin-up and -down electrons in the reservoir. $V_S$ can be computed from $V_S=\mu_\uparrow-\mu_\downarrow$.

Since we are interested in a situation with an external magnetic field, the lowest subband in the wire is split into the spin-up and -down subbands [Fig.~\ref{fig_wire}(b)]. Let us assume that the band bottom $E_\uparrow$ for spin-up electrons is lower than the band bottom $E_\downarrow$ for spin-down electrons. We will assume that the electro-chemical potential $\mu$ of the wire is the same for the spin-up and -down electrons and stays between $E_\uparrow$ and $E_\downarrow$. We will also assume that the temperature
$T\ll E_\downarrow-E_\uparrow$. Then only the lower spin-up subband is populated. Our main focus is on the limit of small $V_S$. By changing the electric potential of the wire, the chemical potential $\mu$
can be made close to $\mu_\uparrow$. In view of the smallness of $V_S$ this implies that the chemical potential $\mu_\downarrow$ of the spin-down electrons in the reservoir is also close to $\mu$. In such situation we expect no tunneling of spin-down electrons between the wire and the reservoir. On the other hand, the tunneling of spin-up electrons disappears only at $\mu=\mu_\uparrow$. Thus, a zero total tunneling current indicates $\mu=\mu_\uparrow$. This gives a way to measure the chemical potential $\mu_\uparrow$ solely from a measurement of an electric current.

In order to find $V_S$ one also needs to measure $\mu_\downarrow$. This can be accomplished in one of two ways. One can use another quantum wire with an opposite $g$-factor. In such wire the order of the spin-split subbands reverses and thus an electrical current measurement yields $\mu_\downarrow$. Alternatively, one can work with the same quantum wire as in the first experiment. By changing the electric potential and the charge density of the wire, one drives the chemical potential $\mu$ so that $\mu$ lies above both band bottoms $E_\uparrow$ and $E_{\downarrow}$. In such situation and focusing on the regime $\mu\approx\mu_\uparrow\approx\mu_\downarrow$, there is a tunneling current of spin-down electrons unless $\mu=\mu_\downarrow$. In that point, the tunneling current is fully spin-polarized. The polarization of the carriers can be determined by a simultaneous measurement of the spin and electric currents between the reservoir and wire. This can be used to deduce $\mu_\downarrow$.



\section{Higher-order cumulants and generating function}
\label{sec:ho}
In this appendix, we briefly describe higher-order cumulants of the heat, electric and spin currents. We consider a general case that the reservoirs have different temperatures, as well as different electric and spin chemical potentials. In general, cumulants can be defined through the generating function, given by
\begin{align}
\mathcal F(&\boldsymbol x, \boldsymbol y, \boldsymbol z, \mathbf T, \mathbf V, \mathbf V_S, \mathbf B) = \nonumber \\
&\lim_{\tau \rightarrow \infty}\frac{1}{\tau} \ln\Bigg\{ \sum_{\Delta \mathbf E, \Delta\mathbf Q, \Delta \mathbf S_z}e^{-(\sum_i x_i \Delta E_i +y_i\Delta Q_{i} + z_i\Delta S_{zi})} \nonumber \\
 & \quad\quad\quad\quad \times P(\Delta \mathbf E, \Delta{\bf Q},\Delta{\bf S}_{z}, \mathbf T, {\bf V}, {\bf V}_S, {\bf B}) \Bigg\}.
\end{align}
where the bold symbols $\boldsymbol x, \boldsymbol y, \boldsymbol z, \mathbf T, \mathbf V, \mathbf V_S$ are vectors, e.g., $\boldsymbol x=(x_1,x_2,\dots, x_N)$, and $N$ is the total number of the reservoirs.

The $n$th order cumulants of heat currents are given by
\begin{align}
&C_{ij\dots k}^h = (\partial_{x_i}-V_i\partial_{y_i} -V_{Si}\partial_{z_i})(\partial_{x_j}-V_j\partial_{y_j} -V_{Sj}\partial_{z_j})\nonumber \\
 &\dots (\partial_{x_k}-V_k\partial_{y_k} -V_{Sk}\partial_{z_k})\mathcal F(\boldsymbol x, \boldsymbol y, \boldsymbol z, \mathbf T, \mathbf V, \mathbf V_S, \mathbf B)
\end{align}
where $\boldsymbol x, \boldsymbol y, \boldsymbol z$ are eventually set to 0 after the derivatives are taken. The number of the indices $i,j,\dots, k$ is $n$, and the indices may or may not be the same. The $n$th order cumulants of electric currents are given by
\begin{equation}
C_{ij\dots k}= \partial_{y_i}\partial_{y_j}\dots\partial_{y_k}\mathcal F(\boldsymbol x, \boldsymbol y, \boldsymbol z, \mathbf T, \mathbf V, \mathbf V_S, \mathbf B)
\end{equation}
and the $n$th order cumulants of spin currents are given by
\begin{equation}
C^S_{ij\dots k}= \partial_{z_i}\partial_{z_j}\dots\partial_{z_k}\mathcal F(\boldsymbol x, \boldsymbol y, \boldsymbol z, \mathbf T, \mathbf V, \mathbf V_S, \mathbf B)
\end{equation}
where again  $\boldsymbol x, \boldsymbol y, \boldsymbol z$ are eventually set to 0 after the derivatives are taken. In general, one can define various cross cumulants of heat, electric and spin currents by certain combinations of the derivatives $(\partial_{x_i}-V_i\partial_{y_i} -V_{Si}\partial_{z_i})$, $\partial_{y_j}$ and $\partial_{z_k}$. Note that the noises and third cumulants in (\ref{15}-\ref{third4}) differ from the above definitions by a conventional factor of 2.

General fluctuation relations can be obtained for higher order cumulants, though they are less relevant to experiments. The fluctuation theorem (\ref{thermo1}) can be translated into a symmetry of the generating function, given by
\begin{align}
\mathcal F(\boldsymbol x, &\boldsymbol y, \boldsymbol z,  \mathbf T, \mathbf V, \mathbf V_S, \mathbf B) = \nonumber\\
&\mathcal F(\boldsymbol \beta -\boldsymbol x, \boldsymbol \xi -\boldsymbol y, \boldsymbol z - \boldsymbol \zeta, \mathbf T, \mathbf V, -\mathbf V_S, -\mathbf B), \label{hc1}
\end{align}
where $\boldsymbol\beta$ is the vector with the entry $\beta_i = 1/T_i$, $\boldsymbol\xi$ is the vector with the entry $\xi_i = - V_i/T_i$, and $\boldsymbol \zeta$ is the vector with the entry $\zeta_i = -V_{Si}/T_i$. To obtain the fluctuation relations, one can perform the Taylor expansion of the generation function $\mathcal F$ on both sides of Eq. (\ref{hc1})  around $\boldsymbol x = \boldsymbol y = \boldsymbol z = \mathbf V = \mathbf V_S =\mathbf 0$ and $T_i = T$. Comparing the Taylor coefficients on the two sides, in principle, one can obtain an infinite number of fluctuation relations. In the main text, we instead use Eq.~(\ref{20}) to derive the fluctuation relations for the lower order cumulants.

\begin{widetext}
\section{Derivation of fluctuation relations from Section IV}
\label{sec:IV}

In this appendix, we provide derivations for some of the relations, listed in Sec.~\ref{sec:example}, from the general expression (\ref{20}).

\subsection{Derivation of two-terminal relations}
\label{app1}
To prove the two-terminal relations (\ref{fdt1})-(\ref{crosstwo4}), we start with a simple helpful identity $\langle \Delta {\bf Q}^{\bf 0} \Delta {\bf S}_{z}^{\bf 0} \rangle=1$. As a consequence,
\begin{equation}
\label{21}
L^{{\bf 0}, {\bf 0}}_{{\boldsymbol \nu},{\boldsymbol \mu}}=\delta_{{\boldsymbol \nu},{\bf 0}}\delta_{{\boldsymbol \mu},{\bf 0}}.
\end{equation}
As mentioned in Sec.~\ref{sec:two_terminal}, in the two-terminal setup, charge conservation makes $\Delta Q_2$ dependent on $\Delta Q_1$ and the gauge invariance principle allows us to set $V_2=0$. So, we can omit the indices $n_2$  and $\nu_2$, and use a shorter notation $L_{\nu_1\mu_1\mu_2}^{n_1m_1m_2}$ instead of the full notation $L^{{\bf n}, {\bf m}}_{{\boldsymbol \nu},{\boldsymbol \mu}}$.

Let us derive (\ref{fdt1}) and (\ref{or1}). Taking the left hand side of (\ref{20}) to be $L^{200}_{000}(\mathbf B), L^{100}_{100}(\mathbf B), L^{000}_{200}(\mathbf B)$ respectively, we have
\begin{align}
L_{000}^{200}(\mathbf B) & = L_{000}^{200}(-\mathbf B), \label{eq48}\\
L_{100}^{100}(\mathbf B) & = -L_{100}^{100}(-\mathbf B)-L_{000}^{200}(-\mathbf B), \nonumber\\
L_{200}^{000}(\mathbf B) & = L_{200}^{000}(-\mathbf B) + L_{100}^{100}(-\mathbf B) + \frac{1}{2}L_{000}^{200}(-\mathbf B). \nonumber
\end{align}
Making use of (\ref{21}), we obtain
\begin{align}
L^{200}_{000}(\mathbf B) & = -2 L^{100}_{100}(\mathbf B), \label{eq51}\\
L^{100}_{100}(\mathbf B)  & = L_{100}^{100}(-\mathbf B) \label{eq52}.
\end{align}
With the definition (\ref{19}) of $L$ and the definitions given in Sec. \ref{sec:definition}, we can identify $L_{000}^{200} = S_{11}/2$, $L_{100}^{100} = -T \partial I_1/\partial V_1$. One immediately obtains (\ref{fdt1}), (\ref{or1}) from (\ref{eq51}), (\ref{eq52}).  Similarly, we obtain (\ref{fdt2}), (\ref{or2}) by taking the left hand side of (\ref{20}) to be $L^{020}_{000}(\mathbf B), L^{010}_{010}(\mathbf B), L^{000}_{020}(\mathbf B)$.

To derive (\ref{fdt3}) and (\ref{or3}), we take the left hand side of (\ref{20}) to be $L^{110}_{000}(\mathbf B)$, $L^{100}_{010}(\mathbf B)$, $L^{010}_{100}(\mathbf B)$ and $L^{000}_{110}(\mathbf B)$ respectively. We then have
\begin{align}
\label{24}
L^{110}_{000}({\bf B})& =-L^{110}_{000}(-{\bf B}),\\
L^{100}_{010}({\bf B})& =L^{100}_{010}(-{\bf B})+ L^{110}_{000}(-{\bf B}), \nonumber\\
L^{010}_{100}({\bf B})& =L^{010}_{100}(-{\bf B})+ L^{110}_{000}(-{\bf B}), \nonumber\\
0=L^{000}_{110}({\bf B}) & =-L^{000}_{110}(-{\bf B})-L^{100}_{010}(-{\bf B}) - L^{010}_{100}(-{\bf B})- L^{110}_{000}(-{\bf B}),\nonumber
\end{align}
and hence
\begin{align}
L^{110}_{000}({\bf B})& =-L^{100}_{010}({\bf B})-L^{010}_{100}({\bf B}).\label{eq53} \\
L^{100}_{010}({\bf B})& =-L^{010}_{100}(-{\bf B}). \label{eq54}
\end{align}
With the identifications $L^{110}_{000}= S^{\rm cr}_{11}/2$, $L^{100}_{010}=-T \partial I_1/\partial V_{S1} $, and $L^{010}_{100} = -T \partial I_{S1}/\partial V_1$, we immediately obtain (\ref{fdt3}), (\ref{or3}) from (\ref{eq53}), (\ref{eq54}).

As an aside, we find that $S_{11}(\mathbf B)= S_{11}(-\mathbf B)$ and $S_{11}^{\rm cr}(\mathbf B) = -S_{11}^{\rm cr}(-\mathbf B)$ from the equations (\ref{eq48}) and (\ref{24}) respectively. A similar relation $S_{12}^S(\mathbf B) = S_{12}^S(-\mathbf B)$ holds.

Next, we derive relations for nonlinear coefficients and the third cumulants. We first prove (\ref{t1}), (\ref{t3}), (\ref{nc1}) and (\ref{crosstwo1}). Taking the left hand side of (\ref{20}) to be $L^{300}_{000}(\mathbf B),L^{200}_{100}(\mathbf B), L^{100}_{200}(\mathbf B), L^{100}_{300}(\mathbf B)$ respectively, we have
\begin{align}
L^{300}_{000}(\mathbf B) &=- L^{300}_{000}(-\mathbf B), \nonumber \\
L^{200}_{100}(\mathbf B) &= L^{200}_{100}(-\mathbf B) + L^{300}_{000}(-\mathbf B), \label{eq55}\\
L^{100}_{200}(\mathbf B) &=- L^{100}_{200}(-\mathbf B) - L^{200}_{100}(-\mathbf B)- \frac{1}{2}L^{300}_{000}(-\mathbf B),\nonumber \\
0=L^{000}_{300}(\mathbf B) &= L^{000}_{300}(-\mathbf B) + L^{100}_{200}(-\mathbf B) + \frac{1}{2} L^{200}_{100}(-\mathbf B)+\frac{1}{6} L^{300}_{000}(-\mathbf B), \nonumber
\end{align}
With the equation (\ref{21}) and with some computation, we arrive at
\begin{align}
L^{300}_{000}(\mathbf B) & =  -3 L^{200}_{100}(\mathbf B) - 6 L^{100}_{200}(\mathbf B), \nonumber\\
L^{200}_{100}(\mathbf B) + L^{200}_{100}(-\mathbf B) & = -2\left[L^{100}_{200}(\mathbf B) + L^{100}_{200}(-\mathbf B)\right], \nonumber\\
L^{200}_{100}(\mathbf B) - L^{200}_{100}(-\mathbf B) & = -6\left[L^{100}_{200}(\mathbf B) - L^{100}_{200}(-\mathbf B)\right], \nonumber\\
2L^{100}_{200}(\mathbf B) & = L^{200}_{100}(-\mathbf B) + 4 L^{100}_{200}(-\mathbf B). \nonumber
\end{align}
Recognizing that $2L^{300}_{000} = -C_{111}$, $2 L^{200}_{100} = T\partial S_{11}/\partial V_1$, and $2 L^{100}_{200} = -T^2\partial^2 I_{1}/\partial V_1^2$, we immediately obtain (\ref{t1}), (\ref{t3}), (\ref{nc1}) and (\ref{crosstwo1}) from (\ref{eq55}) and the above four equations.

To derive (\ref{t2}), (\ref{t4}), (\ref{nc2}) and (\ref{crosstwo3}), we consider the following consequences of (\ref{20})
\begin{align}
L^{030}_{000}(\mathbf B) &= L^{030}_{000}(-\mathbf B),\nonumber \\
L^{020}_{010}(\mathbf B) &=-L^{020}_{010}(-\mathbf B) - L^{030}_{000}(-\mathbf B), \nonumber\\
L^{010}_{020}(\mathbf B) &= L^{010}_{020}(-\mathbf B) + L^{020}_{010}(-\mathbf B) + \frac{1}{2}L^{030}_{000}(-\mathbf B), \nonumber\\
L^{000}_{030}(\mathbf B) &= -L^{000}_{030}(-\mathbf B) - L^{010}_{020}(-\mathbf B) - \frac{1}{2} L^{020}_{010}(-\mathbf B)-\frac{1}{6} L^{030}_{000}(-\mathbf B).\nonumber
\end{align}
Similarly, we are able to obtain
\begin{align}
L^{030}_{000}(\mathbf B) & =  -3 L^{020}_{010}(\mathbf B) - 6 L^{010}_{020}(\mathbf B),\nonumber\\
L^{020}_{010}(\mathbf B) + L^{020}_{010}(-\mathbf B) & = -6\left[L^{010}_{020}(\mathbf B) + L^{010}_{020}(-\mathbf B)\right],\nonumber \\
L^{020}_{010}(\mathbf B) - L^{020}_{010}(-\mathbf B) & = -2\left[L^{010}_{020}(\mathbf B) - L^{010}_{020}(-\mathbf B)\right],\nonumber \\
2L^{010}_{020}(\mathbf B) & = - L^{020}_{010}(-\mathbf B)-4L^{010}_{020}(-\mathbf B).  \nonumber
\end{align}
Then, the relations (\ref{t2}), (\ref{t4}), (\ref{nc2}) and (\ref{crosstwo3}) immediately follow.

Finally, we prove (\ref{crosstwo2}) and (\ref{crosstwo4}). Consider the consequences from (\ref{20})
\begin{align}
L^{100}_{020}(\mathbf B) &= -L^{100}_{020}(-\mathbf B)-L^{110}_{010}(-\mathbf B)-\frac{1}{2}L^{120}_{000}(-\mathbf B), \nonumber\\
L^{000}_{120}(\mathbf B) &=L^{000}_{120}(-\mathbf B)+L^{010}_{110}(-\mathbf B)+\frac{1}{2}L^{020}_{100}(-\mathbf B) +L^{100}_{020}(-\mathbf B)+L^{110}_{010}(-\mathbf B)+\frac{1}{2}L^{120}_{000}(-\mathbf B), \nonumber\\
L^{010}_{200}(\mathbf B) &= L^{010}_{200}(-\mathbf B)+L^{110}_{100}(-\mathbf B)+\frac{1}{2}L^{210}_{000}(-\mathbf B), \nonumber\\
L^{000}_{210}(\mathbf B) &= -L^{000}_{210}(-\mathbf B) - L^{100}_{110}(-\mathbf B)-\frac{1}{2}L^{200}_{010}(-\mathbf B) -L^{010}_{200}(-\mathbf B)-L^{110}_{100}(-\mathbf B)-\frac{1}{2}L^{210}_{000}(-\mathbf B). \nonumber
\end{align}
It follows that
\begin{align}
L^{100}_{020}(\mathbf B) & =L^{010}_{110}(-\mathbf B)+\frac{1}{2}L^{020}_{100}(-\mathbf B), \\
L^{010}_{200}(\mathbf B) & =- L^{100}_{110}(-\mathbf B)-\frac{1}{2}L^{200}_{010}(-\mathbf B).
\end{align}
Recalling that $2L^{100}_{020} = -T^2 \partial^2 I_1/\partial V_{S1}^2$, $L^{010}_{110} = -T^2\partial^2I_{S1}/\partial V_1\partial V_{S1}$, $2L^{020}_{100} = T\partial S_{11}^S/\partial V_1$, $2L^{010}_{200} = - T^2\partial^2 I_{S1}/\partial V_1^2$, $L^{100}_{110}= -T^2\partial I_1/\partial V_1\partial V_{S1}$ and $2L^{200}_{010} = T\partial S_{11}/\partial V_{S1} $, we immediately obtain (\ref{crosstwo2}) and (\ref{crosstwo4}).

\subsection{Derivation of multi-terminal relations}
\label{app2}

The derivations of the multi-terminal relations (\ref{multi1})-(\ref{cross4}) are very similar to the derivations in the two-terminal case. Below we show derivations for some representative relations. The rest can be derived in essentially the same way.

Let us derive the relations (\ref{multifdt}) and (\ref{multior}). Let $\boldsymbol i$ be an integer vector with the $i$-th  entry being 1 and the other entries being 0. The notations $\boldsymbol j, \boldsymbol k, \dots $ are similarly defined. According to the general expression (\ref{20}), we have
\begin{align}
L^{\boldsymbol 0, \boldsymbol j}_{\boldsymbol i, \mathbf 0}(\mathbf B) & = L^{\boldsymbol 0, \boldsymbol j}_{\boldsymbol i, \mathbf 0}(-\mathbf B) + L^{\boldsymbol i, \boldsymbol j}_{\boldsymbol 0, \mathbf 0}(-\mathbf B), \\
L^{\boldsymbol i, \boldsymbol 0}_{\boldsymbol 0, \boldsymbol j}(\mathbf B) & = L^{\boldsymbol i, \boldsymbol 0}_{\boldsymbol 0, \boldsymbol j}(-\mathbf B) + L^{\boldsymbol i, \boldsymbol j}_{\boldsymbol 0, \mathbf 0}(-\mathbf B), \\
L^{\boldsymbol 0, \boldsymbol 0}_{\boldsymbol i, \boldsymbol j}(\mathbf B) & = -L^{\boldsymbol 0, \boldsymbol 0}_{\boldsymbol i, \boldsymbol j}(-\mathbf B) - L^{\boldsymbol i, \boldsymbol 0}_{\boldsymbol 0, \boldsymbol j}(-\mathbf B) - L^{\boldsymbol 0, \boldsymbol j}_{\boldsymbol i, \mathbf 0}(-\mathbf B) -  L^{\boldsymbol i, \boldsymbol j}_{\boldsymbol 0, \mathbf 0}(-\mathbf B).
\end{align}
Using the identity (\ref{21}), we find
\begin{align}
L^{\boldsymbol i, \boldsymbol j}_{\boldsymbol 0, \mathbf 0}(\mathbf B) & = - L^{\boldsymbol i, \boldsymbol 0}_{\boldsymbol 0, \boldsymbol j}(\mathbf B)\nonumber  - L^{\boldsymbol 0, \boldsymbol j}_{\boldsymbol i, \mathbf 0}(\mathbf B), \\
L^{\boldsymbol i, \boldsymbol 0}_{\boldsymbol 0, \boldsymbol j}(\mathbf B) & =  - L^{\boldsymbol 0, \boldsymbol j}_{\boldsymbol i, \boldsymbol 0}(-\mathbf B).
\end{align}
Recall that $L^{\boldsymbol i, \boldsymbol j}_{\boldsymbol 0, \mathbf 0} = S^{\rm cr}_{ij}/2$, $L^{\boldsymbol i, \boldsymbol 0}_{\boldsymbol 0, \boldsymbol j}= -T\partial I_{i}/\partial V_{Sj}$ and $L^{\boldsymbol 0, \boldsymbol j}_{\boldsymbol i, \boldsymbol 0}= -T\partial I_{Sj}/\partial V_i$. We immediately obtain (\ref{multifdt}) and (\ref{multior}).

Let us also derive (\ref{m1}), (\ref{m2}), (\ref{m3}), (\ref{m4}) and (\ref{cross3}). For simplicity, we assume $\boldsymbol i \neq \boldsymbol j \neq \boldsymbol k$. However, the final result does not depend on this assumption. According to the general expression (\ref{20}), we have
\begin{align}
L^{\boldsymbol 0, \boldsymbol i +\boldsymbol j }_{\boldsymbol 0, \boldsymbol k}(\mathbf B) &=  - L^{\boldsymbol 0, \boldsymbol i +\boldsymbol j }_{\boldsymbol 0, \boldsymbol k}(-\mathbf B) -L^{\boldsymbol 0, \boldsymbol i +\boldsymbol j+\boldsymbol k }_{\boldsymbol 0, \boldsymbol 0}(-\mathbf B), \label{eq100}\\
L^{\boldsymbol 0, \boldsymbol i }_{\boldsymbol 0, \boldsymbol j+ \boldsymbol k}(\mathbf B) &=   L^{\boldsymbol 0, \boldsymbol i }_{\boldsymbol 0,\boldsymbol j + \boldsymbol k}(-\mathbf B) +  L^{\boldsymbol 0, \boldsymbol i + \boldsymbol j }_{\boldsymbol 0,\boldsymbol k}(-\mathbf B) + L^{\boldsymbol 0, \boldsymbol i + \boldsymbol k }_{\boldsymbol 0,\boldsymbol j}(-\mathbf B) + L^{\boldsymbol 0, \boldsymbol i +\boldsymbol j+\boldsymbol k }_{\boldsymbol 0, \boldsymbol 0}(-\mathbf B),\nonumber \\
L^{\boldsymbol 0, \boldsymbol 0 }_{\boldsymbol 0, \boldsymbol i+ \boldsymbol j+ \boldsymbol k}(\mathbf B) & = -L^{\boldsymbol 0, \boldsymbol 0 }_{\boldsymbol 0, \boldsymbol i+ \boldsymbol j+ \boldsymbol k}(-\mathbf B)- L^{\boldsymbol 0, \boldsymbol i }_{\boldsymbol 0,\boldsymbol j + \boldsymbol k}(-\mathbf B) - L^{\boldsymbol 0, \boldsymbol j}_{\boldsymbol 0,\boldsymbol k + \boldsymbol i}(-\mathbf B) - L^{\boldsymbol 0, \boldsymbol k }_{\boldsymbol 0,\boldsymbol i + \boldsymbol j}(-\mathbf B)\nonumber\\ &- L^{\boldsymbol 0, \boldsymbol i + \boldsymbol j }_{\boldsymbol 0,\boldsymbol k}(-\mathbf B) -L^{\boldsymbol 0, \boldsymbol j + \boldsymbol k }_{\boldsymbol 0,\boldsymbol i}(-\mathbf B) - L^{\boldsymbol 0, \boldsymbol i + \boldsymbol k }_{\boldsymbol 0,\boldsymbol j}(-\mathbf B) -L^{\boldsymbol 0, \boldsymbol i +\boldsymbol j+\boldsymbol k }_{\boldsymbol 0, \boldsymbol 0}(-\mathbf B). \nonumber
\end{align}
Using the relations, obtained from the above by cyclic permutations of the indices ${\boldsymbol i}, {\boldsymbol j}, {\boldsymbol k}$ and the identity (\ref{21}), it is not hard to show that
\begin{align}
L^{\boldsymbol 0, \boldsymbol i +\boldsymbol j+\boldsymbol k }_{\boldsymbol 0, \boldsymbol 0}(\mathbf B) = - L^{\boldsymbol 0, \boldsymbol i }_{\boldsymbol 0,\boldsymbol j + \boldsymbol k}(\mathbf B) - L^{\boldsymbol 0, \boldsymbol i + \boldsymbol j }_{\boldsymbol 0,\boldsymbol k}(\mathbf B)+ c.p., \nonumber
\end{align}
where $c.p.$ stands for cyclic permutations of the indices  ${\boldsymbol i}, {\boldsymbol j}, {\boldsymbol k}$, and
\begin{align}
\left[L^{\boldsymbol 0, \boldsymbol i + \boldsymbol j }_{\boldsymbol 0,\boldsymbol k}(\mathbf B) + c.p.\right] + \left[L^{\boldsymbol 0, \boldsymbol i + \boldsymbol j }_{\boldsymbol 0,\boldsymbol k}(-\mathbf B) + c.p.\right] = & -3\left\{\left[L^{\boldsymbol 0, \boldsymbol i }_{\boldsymbol 0,\boldsymbol j + \boldsymbol k}(\mathbf B) +c.p.\right] + \left[L^{\boldsymbol 0, \boldsymbol i }_{\boldsymbol 0,\boldsymbol j + \boldsymbol k}(-\mathbf B) + c.p.\right]\right\}, \nonumber\\
\left[L^{\boldsymbol 0, \boldsymbol i + \boldsymbol j }_{\boldsymbol 0,\boldsymbol k}(\mathbf B) + c.p.\right] - \left[L^{\boldsymbol 0, \boldsymbol i + \boldsymbol j }_{\boldsymbol 0,\boldsymbol k}(-\mathbf B) + c.p.\right] = & -\left\{\left[L^{\boldsymbol 0, \boldsymbol i }_{\boldsymbol 0,\boldsymbol j + \boldsymbol k}(\mathbf B) +c.p.\right] - \left[L^{\boldsymbol 0, \boldsymbol i }_{\boldsymbol 0,\boldsymbol j + \boldsymbol k}(-\mathbf B) + c.p.\right]\right\},\nonumber\\
L^{\boldsymbol 0, \boldsymbol i }_{\boldsymbol 0, \boldsymbol j+ \boldsymbol k}(\mathbf B)  = & -L^{\boldsymbol 0, \boldsymbol j + \boldsymbol k }_{\boldsymbol 0,\boldsymbol i}(-\mathbf B)- L^{\boldsymbol 0, \boldsymbol j}_{\boldsymbol 0,\boldsymbol k + \boldsymbol i}(-\mathbf B) - L^{\boldsymbol 0, \boldsymbol k }_{\boldsymbol 0,\boldsymbol i + \boldsymbol j}(-\mathbf B). \nonumber
\end{align}
Recall that $2L^{\boldsymbol 0, \boldsymbol i + \boldsymbol j +\boldsymbol k}_{\boldsymbol 0,\boldsymbol 0} = -C_{ijk}^S$, $2L^{\boldsymbol 0, \boldsymbol i + \boldsymbol j }_{\boldsymbol 0,\boldsymbol k} = T\partial S^S_{ij}/\partial V_{Sk} $ and $L^{\boldsymbol 0, \boldsymbol i }_{\boldsymbol 0,\boldsymbol j + \boldsymbol k} =- T^2\partial^2 I_{Si}/\partial V_{Si}\partial V_{Sj}$. We immediately obtain (\ref{m1}), (\ref{m3}), (\ref{m4}) and (\ref{cross3}). The relation (\ref{m2}) follows from (\ref{eq100}).

Finally, we derive the relation (\ref{cross2}). According to the general expression (\ref{20}), we have
\begin{align}
L^{\boldsymbol k,\boldsymbol 0}_{\boldsymbol 0, \boldsymbol i + \boldsymbol j}(\mathbf B)  &=  -L^{\boldsymbol k,\boldsymbol 0}_{\boldsymbol 0, \boldsymbol i + \boldsymbol j}(-\mathbf B) - L^{\boldsymbol k,\boldsymbol i}_{\boldsymbol 0, \boldsymbol j}(-\mathbf B)-  L^{\boldsymbol k,\boldsymbol j}_{\boldsymbol 0, \boldsymbol i}(-\mathbf B) - L^{\boldsymbol k,\boldsymbol i + \boldsymbol j}_{\boldsymbol 0, \boldsymbol 0}(-\mathbf B), \\
L^{\boldsymbol 0,\boldsymbol 0}_{\boldsymbol k, \boldsymbol i + \boldsymbol j}(\mathbf B) & =   L^{\boldsymbol 0,\boldsymbol 0}_{\boldsymbol k, \boldsymbol i + \boldsymbol j}(-\mathbf B) + L^{\boldsymbol 0,\boldsymbol i}_{\boldsymbol k, \boldsymbol j}(-\mathbf B) +  L^{\boldsymbol 0,\boldsymbol j}_{\boldsymbol k, \boldsymbol i}(-\mathbf B) + L^{\boldsymbol 0,\boldsymbol i + \boldsymbol j}_{\boldsymbol k, \boldsymbol 0}(-\mathbf B) \nonumber\\
& + L^{\boldsymbol k,\boldsymbol 0}_{\boldsymbol 0, \boldsymbol i + \boldsymbol j}(-\mathbf B) + L^{\boldsymbol k,\boldsymbol i}_{\boldsymbol 0, \boldsymbol j}(-\mathbf B) +  L^{\boldsymbol k,\boldsymbol j}_{\boldsymbol 0, \boldsymbol i}(-\mathbf B) + L^{\boldsymbol k,\boldsymbol i + \boldsymbol j}_{\boldsymbol 0, \boldsymbol 0}(-\mathbf B).
\end{align}
where we have assumed $\boldsymbol i \neq \boldsymbol j$ for simplicity, but the final result holds even if  $\boldsymbol i = \boldsymbol j$. Using the identity (\ref{21}) and adding the above two equations, we obtain
\begin{equation}
L^{\boldsymbol k,\boldsymbol 0}_{\boldsymbol 0, \boldsymbol i + \boldsymbol j}(\mathbf B) =L^{\boldsymbol 0,\boldsymbol i + \boldsymbol j}_{\boldsymbol k, \boldsymbol 0}(-\mathbf B)+ L^{\boldsymbol 0,\boldsymbol i}_{\boldsymbol k, \boldsymbol j}(-\mathbf B) + L^{\boldsymbol 0,\boldsymbol j}_{\boldsymbol k, \boldsymbol i}(-\mathbf B).
\end{equation}
Recalling that $2L^{\boldsymbol 0,\boldsymbol i + \boldsymbol j}_{\boldsymbol k, \boldsymbol 0}= T\partial S_{ij}^S/\partial V_k$, $L^{\boldsymbol k,\boldsymbol 0}_{\boldsymbol 0, \boldsymbol i + \boldsymbol j}=-T^2\partial^2 I_{k}/\partial V_{Si}\partial V_{Sj}$ and $ L^{\boldsymbol 0,\boldsymbol j}_{\boldsymbol k, \boldsymbol i}=-T^2\partial^2 I_{Sj}/\partial V_{k}\partial V_{Si}$, we immediately obtain (\ref{cross2}).
\end{widetext}

\begin{figure}[b]
\includegraphics{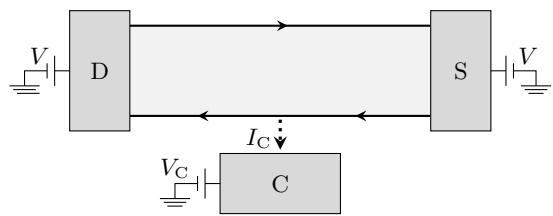}
\caption{Three-terminal quantum Hall setup.}\label{fig_hall}
\end{figure}

\section{Spin transport in a quantum Hall system.}
\label{sec:chiral}

In this appendix we briefly consider an example of a chiral system. In chiral systems there is transport in one direction only, e.g., clockwise.
Chiral transport results in a `stronger' casualty then usual: Not only past does not depend on the future but also what happens upstream does not depend on what happens downstream. Such stronger causality principle greatly simplifies the fluctuation relations and leads to stronger restrictions on transport coefficients than in non-chiral systems \cite{wang1,wang2,wang3}.

Our example is illustrated in Fig.~\ref{fig_hall}. We consider a two-dimensional electron gas in a strong magnetic field in the integer quantum Hall regime with the filling factor $\nu=1$. The bulk of the sample is gapped and all transport occurs on the edges. There is also tunneling into conductor C. The source S and the drain D are maintained at the same temperature $T$ and voltage $V$. There is a voltage and temperature gradient between S and C. We are interested in the spin current $I_C$, flowing into C, and the spin current noises $S_D$ and $S_C$, detected in D and C. The current is fully polarized at $\nu=1$. Thus, the spin current $I_S=-I/2e$. Hence, a non-equilibrium fluctuation-dissipation theorem for spin currents and noises
can be deduced from the results for the noises of the charge currents \cite{wang1,wang2,wang3}:
\begin{equation}
\label{dfdt}
S_D=S_C+\frac{2T}{e}\frac{\partial I_C}{\partial V}+\frac{T}{2\pi}.
\end{equation}
Note that Eq. (\ref{dfdt}) relates  quantities defined at the same magnetic field. Note also that Eq. (\ref{dfdt}) holds in the nonlinear transport regime, i.e., at a {\it finite} voltage drop $V-V_C$.


\begin{thebibliography}{100}

\bibitem{spintronics} I. \u{Z}utic, J. Fabian, and S. Das Sarma, Rev. Mod. Phys. {\bf 76}, 323 (2004).

\bibitem{kynetics} E. M. Lifshitz and L. P. Pitaevsky, {\it Physical Kynetics} (Butterworth-Heinenann, Oxford, 1981).

\bibitem{BK1} G. N. Bochkov and Yu. E. Kuzovlev, Sov. Phys. JETP {\bf 45}, 125 (1977).

\bibitem{BK2} G. N. Bochkov and Yu. E. Kuzovlev, Sov. Phys. JETP {\bf 49}, 543 (1979).

\bibitem{BK3} G. N. Bochkov and Yu. E. Kuzovlev, Physica A {\bf 106}, 443 (1981).

\bibitem{fl0} U. M. B. Marconi, A. Puglisi, L. Rondoni, and A. Vulpiani, Phys. Rep. {\bf 461}, 111 (2008).

\bibitem{fl1} M. Esposito, U. Harbola, and S. Mukamel Rev. Mod. Phys. {\bf 81}, 1665 (2009).

\bibitem{fl2} M. Campisi, P. H\"anggi, and P. Talkner, Rev. Mod. Phys. {\bf 83}, 771 (2011).

\bibitem{see1}  K. Uchida, S. Takahashi, K. Harii, J. Ieda, W. Koshibae, K. Ando, S. Maekawa, and E. Saitoh, Nature (London) {\bf 455}, 778 (2008).

\bibitem{see2} P. Jacquod, R. S. Whitney, J. Meair, and M. B\"uttiker, Phys. Rev. B {\bf 86}, 155118 (2012).

\bibitem{see3} H. Adachi, K.-i. Uchida, E. Saitoh, and S. Maekawa, Rep. Prog. Phys. {\bf 76}, 036501 (2013).

\bibitem{utsumi15} Y. Utsumi and T. Taniguchi, Phys. Rev. Lett. {\bf 114}, 186601 (2015).

\bibitem{name} P. H\"anggi, Helv. Phys. Acta {\bf 51}, 202 (1978).

\bibitem{Cft} D. J. Evans, E. G. D. Cohen, and G. P. Morriss,
Phys. Rev. Lett. {\bf 71}, 2401 (1993); Erratum, {\it ibid.} {\bf 71}, 3616 (1993).

\bibitem{Jft}  C. Jarzynski, Phys. Rev. Lett. {\bf 78}, 2690 (1997).

\bibitem{Crft} G. E. Crooks, Phys. Rev. E {\bf 60}, 2721 (1999).

\bibitem{saito} K. Saito and Y. Utsumi, Phys. Rev. B {\bf 78}, 115429 (2008).

\bibitem{wang1} C. Wang and D. E. Feldman, Phys. Rev. B {\bf 84}, 235315 (2011).

\bibitem{wang2} C. Wang and D. E. Feldman, Phys. Rev. Lett. {\bf 110}, 030602 (2013).

\bibitem{wang3} C. Wang and D. E. Feldman, Int. J. Mod. Phys. B {\bf 28}, 1430003 (2014).

\bibitem{spin-utsumi} Y. Utsumi and H. Imamura, J. Phys.: Conference Series {\bf 200}, 052030 (2010).

\bibitem{lopez12} R. Lopez, J. S. Lim, and D. Sanchez, Phys. Rev. Lett. {\bf 108}, 246603 (2012).

\bibitem{forster08} H. F\"orster and M. B\"uttiker, Phys. Rev. Lett. {\bf 101}, 136805 (2008).

\bibitem{forster09} H. F\"orster and M. B\"uttiker, arXiv:0903.1431.



\bibitem{exp1} S. Nakamura, Y. Yamauchi, M. Hashisaka, K. Chida, K. Kobayashi, T. Ono, R. Leturcq, K. Ensslin, K. Saito, Y. Utsumi, and A. C. Gossard,
Phys. Rev. Lett {\bf 104}, 080602 (2010).

\bibitem{exp2} S. Nakamura, Y. Yamauchi, M. Hashisaka, K. Chida, K. Kobayashi, T. Ono, R. Leturcq, K. Ensslin, K. Saito, Y. Utsumi, and A. C. Gossard,
Phys. Rev. B {\bf 83}, 155431 (2011).

\bibitem{saito09} Y. Utsumi and K. Saito, Phys. Rev. B {\bf 79}, 235311 (2009).

\bibitem{nasb10} K. E. Nagaev, O. S. Ayvazyan, N. Yu. Sergeeva, and M. B\"uttiker, Phys. Rev. Lett. {\bf 105}, 146802 (2010).



\bibitem{Aft} A. Altland, A. De Martino, R. Egger, and B. Narozhny, Phys. Rev. Lett. {\bf 105}, 170601 (2010).

\bibitem{Bft}  A. Altland, A. De Martino, R. Egger, and B. Narozhny, Phys. Rev. B, {\bf 82}, 115323 (2010).

\bibitem{quench} M. A. Cazalilla, Phys. Rev. Lett. {\bf 97}, 156403 (2006).

\bibitem{datta} S. Datta, {\it Electronic Transport in Mesoscopic System} (Cambridge University Press, Cambridge, 1995).

\bibitem{andrieux09} D. Andrieux, P. Gaspard, T. Monnai, and S. Tasaki, New. J. Phys. {\bf 11}, 043014 (2009).

\bibitem{campisi10}M. Campisi, P. Talkner, and P. H¨anggi, Phys. Rev. Lett. {\bf 105}, 140601 (2010).

\bibitem{silicon} A. M. Tyryshkin, S. Tojo, J. J. L. Morton, H. Riemann, N. V. Abrosimov, P. Becker, H.-J. Pohl, T. Schenkel, M. L.W. Thewalt, K. M. Itoh,
and S. A. Lyon, Nature Materials {\bf 11}, 143 (2012).




\bibitem{3moment} Yu. Bomze, G. Gershon, D. Shovkun, L. S. Levitov, and M. Reznikov,
Phys. Rev. Lett. {\bf 95}, 176601 (2005).



\bibitem{rec6} B. Braunecker, D. E. Feldman, and F. Li, Phys. Rev. B {\bf 76}, 085119 (2007).

\bibitem{sptr1} S. Datta and D. Das, Appl. Phys. Lett. {\bf 56}, 665 (1990).

\bibitem{sptr2} J. Wunderlich, B.-G. Park, A. C. Irvine, L. P. Z\^arbo, E. Rozkotov\'a. P. Nemec, V. Nov\'ak, J. Sinova, and T. Jungwirth, Science {\bf 330}, 1801 (2010).

\bibitem{spin-diffusion} N. Bloembergen, Physica {\bf 15}, 386 (1947).



\bibitem{as1} D. Sanchez and M. B\"uttiker, Phys. Rev. Lett. {\bf 93}, 106802 (2004).

\bibitem{as2} B. Spivak and A. Zyuzin, Phys. Rev. Lett. {\bf 93}, 226801 (2004).

\bibitem{n1}G. Gallavotti and E.G.D. Cohen, Phys. Rev. Lett. {\bf 74}, 2694 (1995).

\bibitem{n2} J. Kurchan, J. Phys. A {\bf 31}, 3719 (1998).

\bibitem{n3} J. Kurchan, Les Houches lecture notes, available at arXiv:0901.1271.

\bibitem{n4} D. S. Golubev, Y. Utsumi, M. Marthaler, and G. Sch\"on, Phys. Rev. B {\bf 84}, 075323 (2011).

\bibitem{IS1} S. O. Valenzuela and M. Tinkham, J. Appl. Phys. {\bf 101}, 09B103 (2007).

\bibitem{IS2} K. Ando, H. Nakayama, Y. Kajiwara, D. Kikuchi, K. Sasage, K. Uchida, K. Ikeda, and E. Saitoh, J. Appl. Phys. {\bf 105}, 07C913 (2009).

\bibitem{IS3} J. Wang, B.-F. Zhu, and R.-B. Liu, Phys. Rev. Lett. {\bf 100}, 086603 (2008).


\end{thebibliography}
\end{document}